\documentclass[conference]{IEEEtran}
\IEEEoverridecommandlockouts

\usepackage{hyperref} 
\usepackage{comment}  
\usepackage{cite}
\usepackage{amsmath,amssymb,amsfonts}
\usepackage{algorithmic}
\usepackage{float}
\usepackage{xcolor,colortbl} 
\usepackage{amsmath,amsfonts} %
\usepackage{graphicx}%
\usepackage{textcomp}%
\usepackage{xcolor}%
\usepackage{url}

\usepackage{array}%
\usepackage{pifont}%

\usepackage{amsthm}%
\usepackage{bm}

\usepackage[linesnumbered,ruled,vlined]{algorithm2e}%
\usepackage{setspace}%

\usepackage{multirow}%
\usepackage{siunitx}
\usepackage{footnote}%
\usepackage[utf8]{inputenc}
\usepackage[english]{babel}

\usepackage{blindtext}

\usepackage{dblfloatfix}%

\usepackage{xcolor,colortbl} 
\usepackage{ctable} 
\graphicspath{ {image/} }

\usepackage{pifont}
\newcommand{\xmark}{\ding{55}}

\usepackage{booktabs}
\usepackage{makecell}
\usepackage[normalem]{ulem}
\usepackage{soul}

\def\BibTeX{{\rm B\kern-.05em{\sc i\kern-.025em b}\kern-.08em
    T\kern-.1667em\lower.7ex\hbox{E}\kern-.125emX}}

\newcommand{\mycolor}[1]{\textcolor{blue}{#1}}
    
\begin{document}

\title{
WaLi: Can Pressure Sensors in HVAC Systems Capture Human Speech?
}

\author{\IEEEauthorblockN{Tarikul Islam Tamiti \qquad Biraj Joshi \qquad Rida Hasan \qquad Anomadarshi Barua \\ Department of Cyber Security Engineering, George Mason University, USA}}


\maketitle

\begin{abstract}

Pressure sensors are an integrated component of modern Heating, Ventilation, and Air Conditioning (HVAC) systems. As these pressure sensors operate within the 0–10 Pa range, support high sampling frequencies of 0.5-2 kHz, and are often placed close to human proximity, they can be used to eavesdrop on confidential speech, since human speech has a similar audible range of 0-10 Pa and a bandwidth of 4 kHz for intelligible quality. This paper presents WaLi, which reconstructs intelligible speech from the low-resolution and noisy pressure sensor data with the following technical contributions: (i)  WaLi reconstructs \textit{intelligible speech} from a minimum of 0.5 kHz sampling frequency of pressure sensors, whereas previous work can only detect hot words/phrases. WaLi uses a complex-valued conformer and Complex Global Attention Block (CGAB) to capture inter-phoneme and intra-phoneme dependencies that exist in the low-resolution pressure sensor data. (ii) WaLi handles the transient noise injected from HVAC fans and duct vibrations by reconstructing both the clean magnitude and phase of the missing frequencies of the low-frequency aliased components. We evaluate our attack on practical HVAC systems located in two anonymous industrial facilities. Extensive studies on real-world pressure sensors show an LSD of 1.24 and a NISQA-MOS of 1.78 for 0.5 kHz to 8 kHz upsampling. We believe that such levels of accuracy pose a significant threat when viewed from a privacy perspective that has not been addressed before for pressure sensors. We also provide defenses for the attack.

\end{abstract}


\vspace{-0.3em}
\section{Introduction}
\label{sec:introduction}
\vspace{-0.150em}

The integration of smart sensors into Heating, Ventilation, and Air Conditioning (HVAC) systems has revolutionized building automation and energy management. 
\mycolor{However, such a rich ecosystem of smart sensors is often considered a “double-edged sword" since they can be used to hamper privacy.} 
To understand what level of leakage is appropriate, the key question boils down to: \textit{What types of sensors can leak privacy from today's HVAC systems and how much information can be inferred from a given sensor data in HVACs?} 

To answer the above question, we need to point out that \textit{pressure sensors} are an integrated component of HVAC systems. Pressure sensors often operate within the pressure range of 0-10 Pa and have a high sampling frequency of 0.5-2 kHz \cite{siemens_qbm2030,honeywell_abp2,cfsensor_xgzp6887d}, which is essential for the dynamic control of fans, dampers, and air handling units for real-time monitoring and fast response in today's HVAC systems \cite{ghosh2016multirate,kelman2011model,erickson2009energy}. These pressure sensors are often placed close to humans and installed in room walls, near diffusers, or within ventilation grilles where human occupancy is high to control indoor air quality (IAQ), to monitor thermal comfort, or room-level pressure regulation in spaces like industries, hospitals, or cleanrooms. As these pressure sensors operate within the 0–10 Pa range, support high sampling frequencies of 0.5-2 kHz, and are often placed close to human proximity, they can eavesdrop on confidential speech, since human speech has a similar audible range of 0-10 Pa \cite{kinsler2000fundamentals} and a bandwidth of 4 kHz for intelligible quality.

\mycolor{To eavesdrop on confidential speech with intelligible quality, the attacker must reconstruct speech from pressure sensor data} 
by answering the following two questions: \textit{\textbf{(1)} How to reconstruct intelligible speech of 4 kHz bandwidth from 0.5-2 kHz sampling frequencies of pressure sensors? And \textbf{(2)} \mycolor{How do we handle transient noise injected from HVAC fans, duct vibrations, physical shocks, or turbulent airflow while reconstructing intelligible speech from pressure sensor data?}}

In this paper, we answer the above two questions by proposing WaLi (\textbf{\underline{Wa}}ll can \textbf{\underline{Li}}sten), which can reconstruct intelligible speech with a bandwidth up to 4 kHz from a lowest sampling frequency of 500 Hz of pressure sensor in transient noisy conditions. WaLi employs the following two technical strategies to answer the above two questions:

\textbf{Strategy 1.} With a sampling rate of 0.5-2 kHz, high-frequency speech components are severely aliased and truncated in pressure sensors, while a few lower pitch frequencies are preserved. Unfortunately, a few low pitches are not sufficient to provide perfect intelligibility \cite{fant1960acoustic} as most of the formants at high frequencies will be missing from the bandwidth. To reconstruct intelligible speech from the severely aliased spectrogram of pressure sensors, the missing high frequencies should be reconstructed from the low-frequency aliased components. To make this happen, WaLi employs conformers \cite{gulati2020conformer}, combining the strengths of Convolutional Neural Networks (CNNs) and Transformers to capture local and global dependencies between low-frequency pitches and missing high frequencies. In addition to conformers, we design \textit{Complex Global Attention Block (CGAB)} to capture the long-range inter-phoneme correlations that exist among pitches and harmonics along both the time and the frequency axes in a spectrogram. Prior work \cite{wang2024vibspeech} published in USENIX only captures correlations along the time axis.

\textbf{Strategy 2.} To prevent the transient noise of HVAC from impacting speech reconstruction, WaLi reconstructs both the clean magnitude and phase of the missing frequencies from the low-frequency aliased components. To jointly reconstruct clean magnitude and phase, WaLi is designed as a \textit{complex-valued network}, which can enhance both the amplitude and phase of pressure sensor data using complex-valued time-frequency (T-F) spectrogram \cite{williamson2016complex}. As complex-valued T-F spectrograms have both magnitude and phase information, WaLi takes the complex-valued T-F spectrograms as input and removes noisy phases from them, motivated by the fact that phase plays a crucial role in speech enhancement \cite{yin2020phasen}. Prior works \cite{wang2024vibspeech,hu2022accear,zhang2023spy} only use real-valued T-F spectrograms and cannot reconstruct a clean phase under noisy conditions. Moreover, WaLi employs a \textit{complex multi-resolution Short-Time Fourier Transformation (STFT) loss} to reconstruct clean magnitude and phase from the noisy pressure sensor data.

We extensively evaluate the effectiveness of WaLi in \ul{two real-world anonymous industrial facilities, using {\color{blue}{six}} evaluation metrics - LSD, NISQA-MOS, PESQ, STOI, {\color{blue}{WER}}, and SI-SDR} (see Section \ref{subsec:Evaluation_Metrics} for the full form of these terms). Our findings highlight the serious privacy implications
of pressure sensors in HVAC systems. 
The implications could be particularly severe in critical environments, such as industrial facilities, corporate offices, and healthcare institutions, where eavesdropping can expose sensitive private information. Our main contributions are summarized as follows:

\textbf{1.} We propose WaLi, an acoustic eavesdropping system that uses pressure sensor data to reconstruct intelligible user speech. \textit{To the best of our knowledge, WaLi is the first method that recovers intelligible speech with an unconstrained vocabulary rather than recognizing individual hot words/phrases from the pressure sensor data.}

\textbf{2.} We use complex-valued architecture to jointly reconstruct both the clean magnitude and phase from the noisy pressure sensor data and extensively evaluate the effectiveness of WaLi using \mycolor{six evaluation} metrics - LSD, NISQA-MOS, PESQ, STOI, \mycolor{WER,} and SI-SDR. A real-world demonstration of the reconstructed audio is provided in the following link: 
{\href{https://sites.google.com/view/pressuresensorwali/home}{\textcolor{blue}{{WaLi}}}.

\vspace{-0.3em}
\section{Background}
\label{sec:background}
\vspace{-0.2em}

\subsection{Human Voice's Pressure Range and Intelligible Bandwidth}
\label{subsecc:Human Voice and Its Pressure Range}
\vspace{-0.2em}

The human voice creates fluctuations in sound pressure, measured in Pascals (Pa). \mycolor{A comparative study of the voice pressure range, given in Table \ref{table:voice SPL} of Appendix \ref{append:Speech Production}, indicates that pressure of speech signals is typically restricted between 0-10 Pa for a wide range of conditions \cite{ccohs2019soundpressure}.} Therefore, to successfully eavesdrop on speech, pressure sensors should also need to be sensitive in the same range of 0-10 Pa. Moreover, for speech intelligibility, only 8 kHz sampling is sufficient, as up to 4 kHz bandwidth is enough for intelligible speech. 

\begin{figure}[h]
\vspace{-0.97320em}
\centering
\includegraphics[width=0.47\textwidth,height=0.11\textheight]{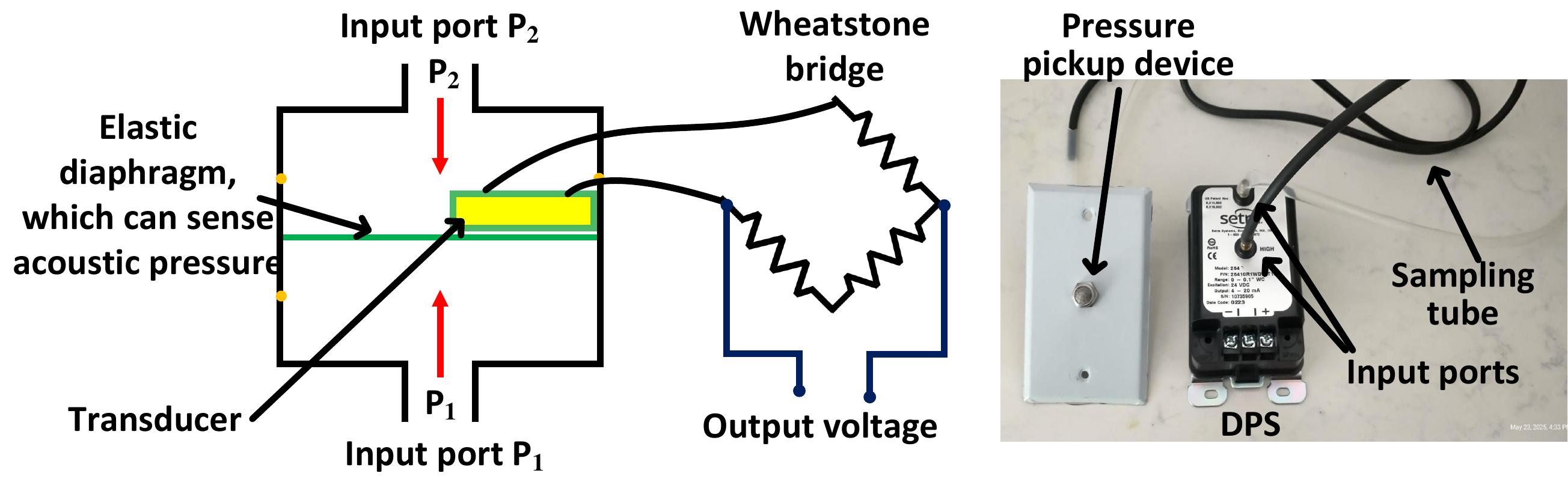}
\vspace{-01.1920em}
\caption{ (Left) Internals of a DPS. (Right) The accessories connected with DPSs (part\# SETRA264) \cite{setra264_datasheet} in a real-world HVAC system.}
\label{fig:DPS_NPR}
\vspace{-1.200em}
\end{figure}

\vspace{-0.2em}
\subsection{Pressure Sensor in HVACs}
\label{subsecc:Pressure Sensor and Its Architecture}
\vspace{-0.2em}

Differential pressure sensors (DPSs) are the state-of-the-art sensors for HVAC systems due to their accurate measurement and reliable operations \cite{gourab2021iot}. Hence, in the rest of the paper, we use the DPS and the pressure sensor terms interchangeably. 

DPSs have an elastic diaphragm 
placed between two pressure input ports $P_1$ and $P_2$ (see Fig. \ref{fig:DPS_NPR} (Left)).  The diaphragm senses the differential pressure $P_1$ - $P_2$ applied to the input ports by changing its shape. The change in the shape of the diaphragm is converted to a proportional output voltage by using a transducer and a Wheatstone bridge \cite{omega_pressure_sensor}. \textit{The diaphragm is sensitive to acoustic pressure and can pick up sound pressure when someone speaks.} In real-world HVAC systems, like part\# SETRA264 \cite{setra264_datasheet}, sampling tubes are connected with either one or both input ports of DPSs. The other end of the sampling tube is connected to pressure pickup devices (see Fig. \ref{fig:DPS_NPR} (Right)), which senses pressure in the environment.   


\begin{table}[ht!]
\vspace{-01.200em}
\footnotesize
    \centering
     \renewcommand{\arraystretch}{0.65}
    \caption{Human audible pressure range and sampling frequencies of pressure sensors in HVACs.}
    \vspace{-0.7800em}
   \begin{tabular}{|p{1.6cm}|p{1.05cm}|p{1.3cm}|p{2.8cm}|}
    \hline
        \cellcolor [gray]{0.85}\textbf{Application} & \cellcolor [gray]{0.85}\textbf{Pressure Range} & \cellcolor [gray]{0.85}\textbf{Sampling Frequency} & \cellcolor [gray]{0.85}\textbf{Function}\\ 
        \hline
        \hline
         Pressure Balancing \cite{guardian} & 0–50 Pa & $\sim$0.5 kHz & Stabilize pressure in connected indoor spaces \\
    \hline
       Air Filter Monitor \cite{superior_hv_series} & 0–150 Pa & $\sim$0.7 kHz & Detect pressure drop to signal filter clogging \\
    \hline
    Duct Static Pressure \cite{sensirion_sdp1108} & 0–200 Pa & $\sim$1 kHz & Ensure optimal airflow and energy use \\
    \hline
    VAV Control \cite{sensirion_sdp1108} & 0–200 Pa & $\sim$2 kHz & Adjust airflow with occupancy or thermal need \\
    \hline
    
    \end{tabular}

    \vspace{-01.20em}
    \label{table:pressureHVACs}
\end{table}

 \vspace{-0.2em}
\subsection{Pressure Range and Sampling Frequencies of DPSs}
\label{subsecc:Pressure Range and Sampling Frequencies of Pressure Sensors}
 \vspace{-0.2em}

Sections \ref{subsecc:Human Voice and Its Pressure Range} points out that for eavesdropping using DPSs in HVACs, the DPSs should be sensitive within 0-10 Pa and should have a sampling frequency close to 8 kHz. In HVAC systems, DPSs operating within the 0–10 Pa range and supporting high sampling frequencies within 0.5-2 kHz are typically used to achieve precise environmental control 
where small pressure changes must be detected accurately and quickly.  \textit{High-frequency sampling enables real-time responsiveness to transient pressure fluctuations, which is essential for dynamic control of fans, dampers, and air handling units.} Common use cases include air filter monitoring \cite{superior_hv_series}, cleanroom pressure regulation \cite{cleanroomsensor}, static duct pressure control, and variable air volume (VAV) management \cite{sensirion_sdp1108}. 
A summary of the pressure sensor range and sampling frequencies in HVACs is given in Table \ref{table:pressureHVACs}, which shows that DPSs in HVACs are sensitive to the audible pressure range of 0-10 Pa.

\begin{figure}[h]
\vspace{-01.10em}
\centering
\includegraphics[width=0.48\textwidth,height=0.11\textheight]{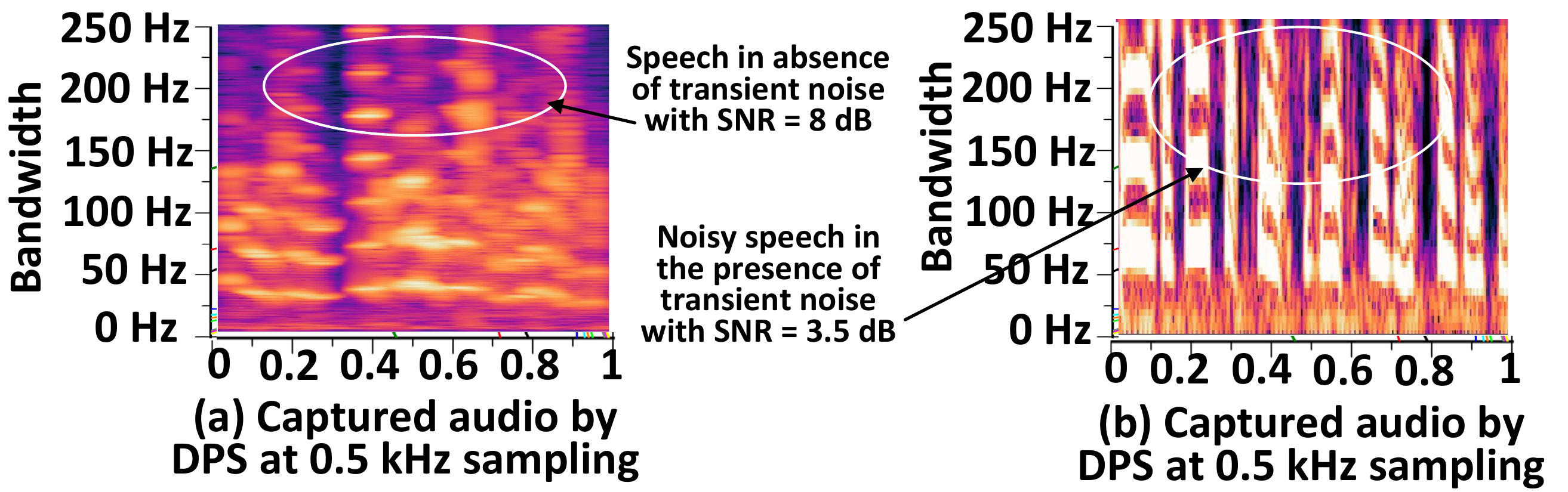}
\vspace{-1.0em}
\caption{(Left) Clean speech captured by SDP 1108 in the absence of noise. (Right) Noisy speech captured by SDP 1108 in the presence of transient noise.}
\label{fig:noisespeech}
\vspace{-0.7500em}
\end{figure}

\begin{figure*}[h]
\vspace{-00.70em}
\centering
\includegraphics[width=0.96\textwidth,height=0.14\textheight]{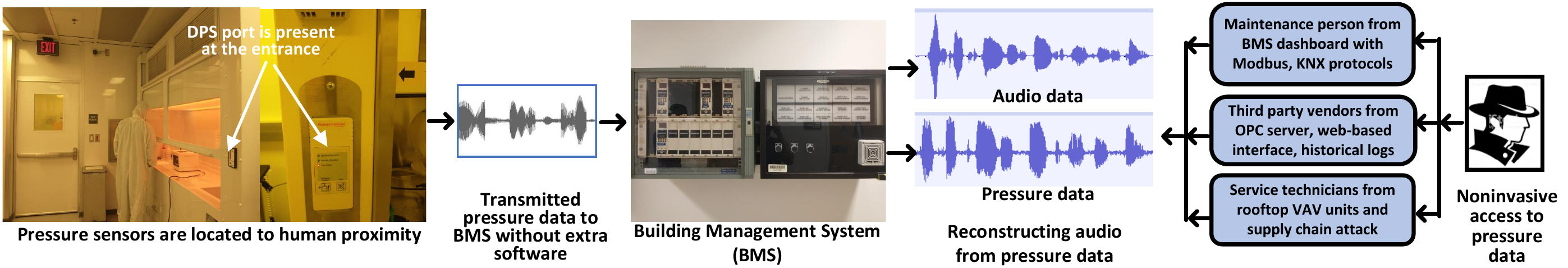}
\vspace{-1.10em}
\caption{(Left) A brief overview of the threat model - WaLi. The victim is unknowingly talking close to the pressure pickup device.}
\label{fig:attack_model}
\vspace{-1.200em}
\end{figure*}

\vspace{-0.7300em}
\subsection{Transient Noise and Importance of Phase Reconstruction}
\label{subsecc:Phase Reconstruction for Intelligible Speech}
\vspace{-0.200em}

The presence of transient noise sources, such as HVAC fans, duct vibrations, physical shocks, or turbulent airflow, injects substantial noise into DPSs in the form of pressure spikes and fluctuating readings \cite{broner1999response}. The impact of noise becomes particularly significant within a low-pressure range of 0–10 Pa and at high sampling frequencies of 0.5-2 kHz. The noisy readings from DPSs impact the intelligible speech reconstruction for successful eavesdropping. Although earlier speech reconstruction methods \cite{hu2022accear, wang2024vibspeech, achamyeleh2024fly} focus on estimating the magnitude spectrum of missing higher frequencies, the phase spectrum is equally important \cite{yin2020phasen} to produce natural and intelligible audio, particularly under transient noisy conditions in real-world HVAC systems. Fig. \ref{fig:noisespeech} shows the impact of noise on pressure sensor data at 0.5 kHz sampling frequency. Transient noise corrupts the phase and magnitude of the speech signal collected by a DPS (SDP 1108 \cite{sensirion_sdp1108}) and deteriorates the Signal-to-Noise Ratio (SNR) from 8 dB to 3.5 dB.  

\vspace{-0.2em}
\section{Threat Model} 
\label{sec:Threat Model}
\vspace{-0.1em}

We discuss the threat model of WaLi below (see Fig. \ref{fig:attack_model}).

\textbf{a) Proximity to sound sources and humans:} 
\mycolor{\textit{To prove that DPSs are often located close to humans, we have evaluated two anonymous facilities - one is an industrial facility and the other is an FDA-compliant cleanroom (see Fig. \ref{fig:DPSsclosetohuman}).}} 


\begin{figure}[h]
\vspace{-0.80em}
\centering
\includegraphics[width=0.48\textwidth,height=0.14\textheight]{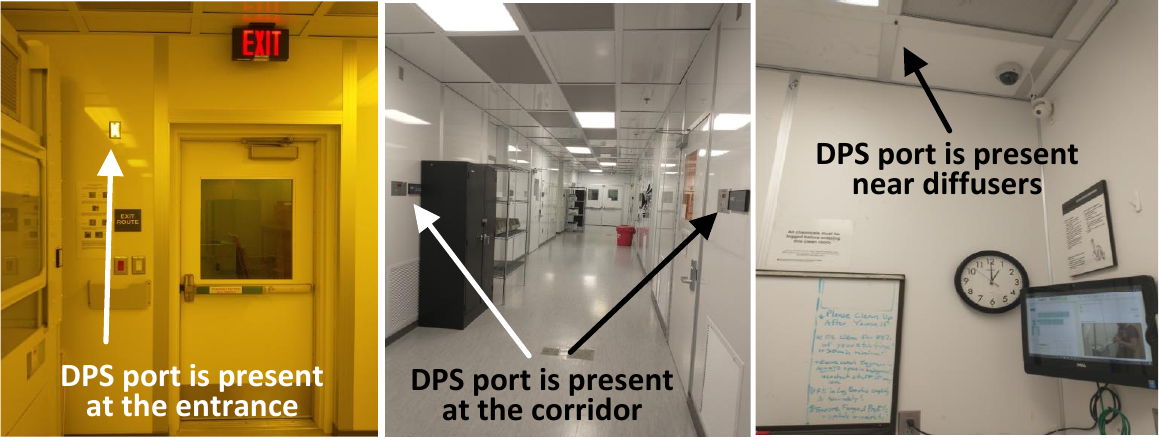}
\vspace{-1.20em}
\caption{(Left \& Middle) Pressure ports are located at hallway entrance and corridor of a cleanroom, and (Right) inside rooms of an industrial facility.}
\label{fig:DPSsclosetohuman}
\vspace{-0.600em}
\end{figure}

\textbf{Fig. \ref{fig:DPSsclosetohuman} supports the fact that DPSs are often found in room entrances, corridors, inside rooms near diffusers, etc., in practical HVAC systems in human proximity and they operate in low-pressure ranges (e.g., 0–250 Pa)}. 
Therefore, they can be a source of eavesdropping. \textit{The main challenges are that DPSs are noisy and use a sampling frequency of 0.5-2 kHz (refer to Table \ref{table:pressureHVACs}). Therefore, we propose WaLi to recover full intelligible speech from the aliased pressure sensor data by reconstructing both amplitude and phase of the speech signal.}


\textbf{b) Attacker's goal:} The attacker can eavesdrop on speech with \textit{unrestricted vocabulary} using HVAC systems where DPSs are located close to human occupancy. In addition to eavesdropping a natural speech, the attacker may also understand how many people are present inside a room/facility, how many of them are males and females, what types of sound sources are present, and from the sound sources, \mycolor{the attacker can also identify what types of activities are going on inside.} 

\textbf{c) Attacker's access level:} Prior works 
require installation of malicious apps to access inertial sensors \cite{wang2024vibspeech, hu2022accear} to collect data for eavesdropping. 
\ul{In contrast, our attack model exploits the normal behavior of HVACs without installing additional software on HVACs.} Access to collect pressure data from DPSs is possible in the following scenarios.

\textbf{First}, an attacker disguised as a malicious employee or maintenance person can access pressure data from the Building Management System (BMS) software dashboard, as in modern buildings, pressure sensors are integrated into the BMS using standard protocols, such as Modbus TCP, and KNX \cite{digikey2023networking}. 

\textbf{Second}, in many cases, the BMS is handled by third-party contractors, or system integrators, especially in commercial buildings, hospitals, labs, and large campuses. 
\ul{In many cases, authorities often outsource teams to provide continuous support and alert handling.} An attacker disguised as one of these third-party vendors can access sensitive pressure data via a web-based interface, historical logs, or an Open Platform Communication (OPC) server.

\textbf{Third}, some commercial HVAC units (e.g., rooftop units, air handling units, variable air volume units) have onboard controllers and internal sensors. Service technicians often access these units for diagnostics in case of failure and troubleshooting. An attacker disguised as one of the technicians can access the pressure data through the control panel and diagnostic port via network protocols.



\textbf{d) Non-invasive attack:} Our eavesdropping attack using the HVAC is non-invasive as it is performed without making direct physical contact with the target DPS. 
However, we expect that attackers can examine the behavior of a similar sensor subjected to acoustic impacts before initiating an actual attack. 

\textbf{e) Attacker's resources and cost:} We assume that the attacker knows how the HVAC system works and how the data can be collected from pressure sensors.

\textbf{f) Assumption:} \ul{Considering the practicality of the attack, we assume that the attacker has sufficient pressure sensor data collected from the victim and does not have matching ground truth audio from the victim. WaLi can reconstruct the audio even if the speaker is different. This makes our attack model flexible compared to prior works} \cite{hu2022accear,achamyeleh2024fly}, \ul{where prior works assume that clean ground-truth speech is available via a microphone or a phishing call.}

\begin{figure*}[htbp]
  \centering
\includegraphics[width=0.89\textwidth,height=0.22\textheight]{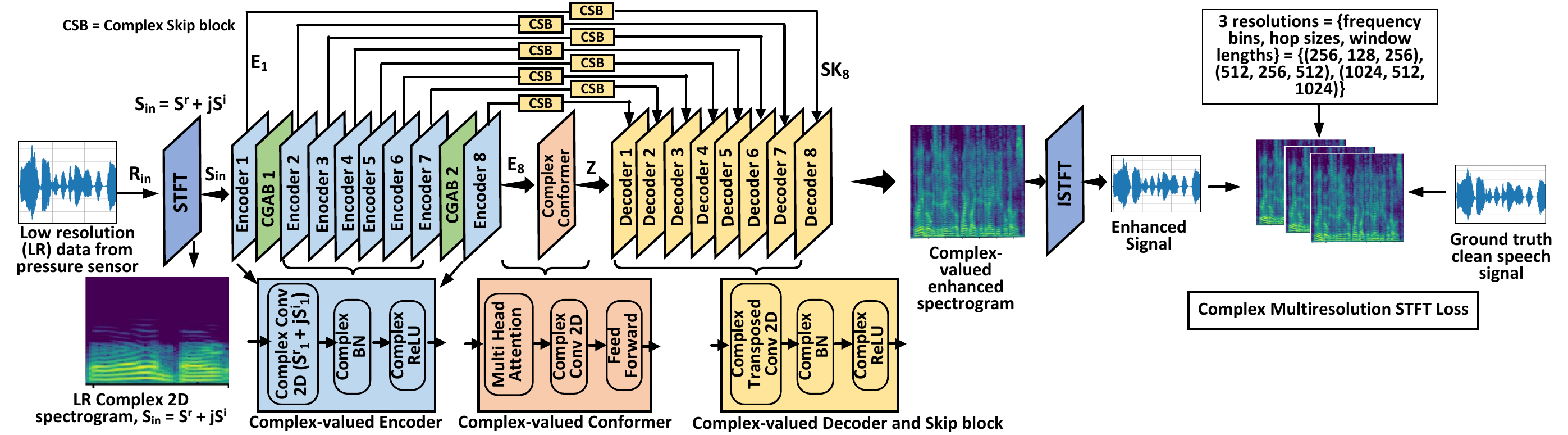}
\vspace{-0.850em}
  \caption{WaLi has complex-valued encoders, decoders, complex-valued skip blocks, CGAB, and complex multiresolution STFT loss. }
  \label{fig:overall_architecture}
  \vspace{-01.59750em}
\end{figure*}

\vspace{-0.20em}
\section{WaLi Architecture Design}
\label{sec:WaLi ARCHITECTURE DESIGN}
\vspace{-0.10em}

WaLi is designed to achieve the following objectives:

\vspace{0.30em}
\noindent
\textbf{a)} WaLi processes the pressure sensor data in the \textit{complex-valued T-F spectrum} and enhances both the amplitude and phase of the pressure  data to reconstruct intelligible speech.

\vspace{0.0em}
\noindent
\textbf{b)} WaLi handles transient interference in 
pressure sensor data using \textit{joint magnitude and phase reconstruction} using a \mycolor{complex-valued T-F spectrogram and network.} 



\textbf{Why the complex-valued spectrogram and network:} A complex-valued T-F spectrogram of a signal contains both magnitude and phase information, not just the energy or amplitude as in a standard real-valued spectrogram. A complex-valued T-F spectrogram is typically generated by the Short-Time Fourier Transform (STFT) and then outputs a complex value $S(t,f)$ at each time-frequency bin, following Eq. \ref{eqn:complex_spec}.

\vspace{-0.65em}
\begin{equation}
\begin{aligned}
S(t,f) = A(t,f) * e^{j \phi (t,f)}
\end{aligned}
\label{eqn:complex_spec}
\vspace{-0.20em}
\end{equation}

where $A(t,f)$ is the magnitude and $\phi(t,f)$ is the phase of the spectrum.  Prior works \cite{hu2022accear, wang2024vibspeech} only use real-valued T-F spectrograms. Therefore, they cannot predict phase and hence, need a vocoder \cite{liu2022neural} to generate audio from real-valued spectrograms. Moreover, real-valued T-F spectrogram-based methods typically use Mean Squared Error (MSE) loss for magnitudes and cannot use phase-reconstruction loss functions \cite{lu2024towards} to improve the phase quality while reconstructing speech from the low-resolution pressure sensor data. Therefore, motivated by the fact that phase plays a crucial role in speech enhancement \cite{yin2020phasen}, WaLi adopts a joint reconstruction of magnitudes and phases. mycolor{As complex-valued spectrograms have both magnitude and phase information (refer to Eqn. \ref{eqn:complex_spec}), WaLi takes complex-valued T-F spectrograms as input and generates complex-valued T-F spectrograms at its output.}  

\textbf{WaLi architecture:} \mycolor{To process the complex-valued T-F spectrograms, we design WaLi as a complex-valued network having a U-Net at its backbone (see Fig. \ref{fig:overall_architecture}) 
with four main components: \textbf{(i)} 16 full complex-valued encoder-decoder blocks} (i.e., 8 encoders and 8 decoders), \textbf{(ii)} complex-valued skip blocks, \textbf{(iii)} complex-valued conformer in the bottleneck layer, and \textbf{(iv)} complex-valued global attention blocks - CGAB. The complex domain processing by WaLi has the potential to recover clean phases from the noisy pressure sensor and to reconstruct magnitudes from the aliased spectrum (i.e., pressure is sampled at a minimum of 0.5 kHz).

\vspace{-0.5em}
\subsection{Complex-Valued Encoders}
\vspace{-0.2em}
\label{subsec:complex_encoder}


\mycolor{The pressure sensor data is transformed to complex-valued T-F spectrogram by STFT and given as} input to the first complex encoder block (E1). Formally, the input low-resolution pressure data $R_{in}$ is first transformed into STFT spectrogram, denoted by $S_{in}$ in Fig. \ref{fig:overall_architecture}. Here, $S_{in} (= S^r+ jS^i) \in \mathbb{C}^{F \times T}$ is a complex-valued spectrogram, where $F$ denotes the number of frequency bins and $T$ denotes the number of time frames.

WaLi adopts \textit{complex-valued encoders}, which is built upon complex-valued convolution to ensure successive extraction of both magnitude and phase from the complex-valued T-F spectrogram. Complex convolution is the key difference between a complex-valued encoder and a real-valued encoder. Formally, the complex-valued T-F spectrogram $S_{in}$ is fed into 2D complex convolution layers of the first encoder to produce feature $S_1 \in \mathbb{C}^{F \times T \times C}$, where C is the number of channels. If a complex kernel is denoted by $W = {W}_r + j{W}_i$, the complex convolution is defined by Eqn. \ref{eqn:complex_conv}.

\vspace{-0.5em}
\begin{equation}
\begin{aligned}
{S}^r_1 &= {W}_r * {S}^r_{in} - {W}_i * {S}^i_{in} + {b}_r, \\
{S}^i_1 &= {W}_r * {S}^i_{in} - {W}_i * {S}^r_{in} + {b}_i,
\end{aligned}
\label{eqn:complex_conv}
\vspace{-0.0em}
\end{equation}



where $*$ denotes the convolution, $S^r_1$ $\&$ $S^i_1$ are real and imaginary parts of  $S_1$, and ${b}_r$ $\&$ ${b}_i$ are bias terms. 

The convolution output is then normalized using Complex Batch Normalization (CBN) for stable training. CBN is an extension of traditional batch normalization to complex-valued neural networks, where inputs, weights, or features \mycolor{have both real and imaginary parts.} Instead of treating real and imaginary parts independently, CBN treats them as a 2D vector and normalizes using a 2×2 covariance matrix using Eqn. \ref{eqn:complex_covariance}.

\vspace{-0.5em}
\begin{equation}
\Sigma_{S_1} = 
\begin{bmatrix}
\text{Var}(S^r_1) & \text{Cov}(S^r_1, S^i_1) \\
\text{Cov}(S^i_1, S^r_1) & \text{Var}(S^i_1)
\end{bmatrix}
\label{eqn:complex_covariance}
\end{equation}
\vspace{-0.95em}

where $\Sigma_{S_1}$ is the covariance matrix of $S_1$. Then the CBN is calculated by Eqn. \ref{eqn:complex_BN}.

\vspace{-0.75em}
\begin{equation}
\text{CBN} = \Sigma_{S_1}^{-\frac{1}{2}} (S_1 - \mu)
\label{eqn:complex_BN}
\end{equation}
\vspace{-01.5em}

where $\Sigma_{S_1}^{-\frac{1}{2}}$ is the inverse square root matrix of $\Sigma_{S_1}$ used to decorrelate and normalize the input $S_1$, and $\mu$ is the complex mean vector of $S_1$. Next, the output from CBN passes through a complex ReLU activation for adding nonlinearity. Formally, first encoder output, denoted by $E_1$ is expressed by Eqn. \ref{eqn:complex_encoder}.

\vspace{-0.70em}
\begin{equation}
E_1 = \text{Complex ReLU(CBN} (S^r_1 + j S^i_1))
\label{eqn:complex_encoder}
\end{equation}
\vspace{-01.65em}

The output of the 1st encoder is given as input to the 2nd encoder, and 2nd encoder's output to third encoder, and so on.  Each complex encoder has a similar complex 2D convolution (Eqn. \ref{eqn:complex_conv}), CBN (Eqn. \ref{eqn:complex_BN}), and complex ReLU (Eqn. \ref{eqn:complex_encoder}).

\vspace{-0.2975em}
\subsection{Complex-Valued Conformer in Bottlenecks}
\label{subsec:conformer_symbolic}
\vspace{-0.1em}


We design complex-valued conformers in the bottleneck layer because 
\mycolor{conformers \cite{gulati2020conformer} combine the strengths of CNNs and Transformers by stacking self-attention and convolutional modules in each layer, resulting in a highly expressive and efficient model for sequential input.} Our complex-valued conformer optimally balances global context with fine-grained local information in phase and amplitude domains for the successful reconstruction of audio from pressure data.

The output of the last (eighth) encoder block, expressed as $E_8 \in \mathbb{R}^{B \times C \times F \times T}$, is fed as input into the conformer. Here,  \(B\) denotes the batch size, \(C\) the number of channels, \(F\) the spectral dimension, and \(T\) the temporal dimension. We rearrange $E_8$ using a permutation and reshaping so that the conformer processes each channel independently. For a given channel, the complex-valued conformer comprises complex multi head self-attention (MHSA), complex feed-forward (FF), and complex convolutional modules. The complex MHSA uses complex-valued queries, keys, and values over the real and imaginary parts. The complex FF module uses complex linear layers, complex ReLU activation, and CBN.

\vspace{-0.3em}
\subsection{Complex-Valued Decoder}
\label{subsec:Complex-Valued Decoder}
\vspace{-0.1em}

The output from the bottleneck layer, denoted by $\mathbf{Z}$ (i.e., latent space), is given as input to the first decoder. The decoder reconstructs the T-F representation from the latent space by employing a complex transposed convolution (upsampling by a factor of two) at each decoder. For a latent complex tensor $\mathbf{Z} = \mathbf{Z}_r + j\,\mathbf{Z}_i,$ the transposed convolution is formulated as:
\begin{equation}
\begin{aligned}
\tilde{\mathbf{Y}}_1^r &= \mathbf{W}_r^T * \mathbf{Z}_r - \mathbf{W}_i^T * \mathbf{Z}_i + \mathbf{b}_r, \\
\tilde{\mathbf{Y}}_1^i &= \mathbf{W}_r^T * \mathbf{Z}_i + \mathbf{W}_i^T * \mathbf{Z}_r + \mathbf{b}_i,
\end{aligned}
\label{eq:complex_transconv}
\end{equation}

where $\tilde{\mathbf{Y}}_1^r$ $\&$ $\tilde{\mathbf{Y}}_1^i$ are real and imaginary parts of  $\tilde{\mathbf{Y}}_1$ (i.e., output after the transposed convolution in the first decoder), and ${b}_r$ $\&$ ${b}_i$ are bias terms. $\tilde{\mathbf{Y}}_1$ is then normalized by CBN and activated by a complex ReLU activation as Eqn. \ref{eq:decoder_activation}.

\vspace{-0.75em}
\begin{equation}
D_1 = \text{Complex ReLU}\Big( \text{CBN}\big(\tilde{\mathbf{Y}}_1^r + j\,\tilde{\mathbf{Y}}_1^i\big) \Big)
\label{eq:decoder_activation}
\end{equation}
\vspace{-01.51em}

where $D_1$ is the output of the first decoder, which is given as input to the 2nd decoder, and so on.  Every complex decoder has a similar complex 2D convolution (Eqn. \ref{eq:complex_transconv}), CBN (Eqn. \ref{eqn:complex_BN}), and complex ReLU activation (Eqn. \ref{eq:decoder_activation}).

\vspace{-0.3em}
\subsection{Complex Skip Block} 
\label{subsec:complexskipblock}
\vspace{-0.13em}

\mycolor{WaLi implements skip blocks in complex domains, inspired by \cite{kothapally2020skipconvnet}, to enable the proper flow of complex-valued high-dimensional features from the encoder's output to appropriate decoders.} Each complex skip block applies a complex convolution on the encoder output,  followed by a CBN and a complex ReLU activation. Formally, the complex skip block's output, denoted by $SK_n$ is (i.e, n is 1 to 8): 


\vspace{-0.95em}
\begin{equation}
SK_n =   \text{Complex ReLU(CBN (Complex Conv} (E_n)))
\label{eq:skipblock}
\end{equation}
\vspace{-01.35em}

where  $\text{Complex Conv}$ is implemented following Eqn. \ref{eqn:complex_conv}, and $E_n$ is the output from the nth encoder (i.e, n is 1 to 8).

\vspace{-0.2em}
\subsection{Complex Global Attention Block (CGAB)}
\label{subsec:CGAB}
\vspace{-0.0em}

Long-range correlations exist along both the time and the frequency axes in a complex-valued T-F spectrogram. As the audio signal embedded in the pressure sensor data is a time series signal, inter-phoneme correlations exist along the time axis. Moreover,  harmonic correlations also exist among pitch and formants along the frequency axis. \textit{As a convolution kernel is limited by its receptive field, standard convolutions cannot capture global correlations on the time and frequency axes in a complex-valued T-F spectrogram.} Therefore, we propose a Complex Global Attention Block (CGAB) to capture long-range correlation from both the T-F axes.  Please note that Frequency Transformation Blocks used in \cite{wang2024vibspeech}, published in USENIX, don't work along the T-F axis (i.e., only time axis). 

CGAB provides attention to the T-F axes of a complex-valued T-F spectrogram by following two steps (see Fig. \ref{fig:CGAB}):

\vspace{-0.8505em}
\begin{figure}[ht!]
  \centering
 \includegraphics[width=0.44\textwidth,height=0.26\textheight]{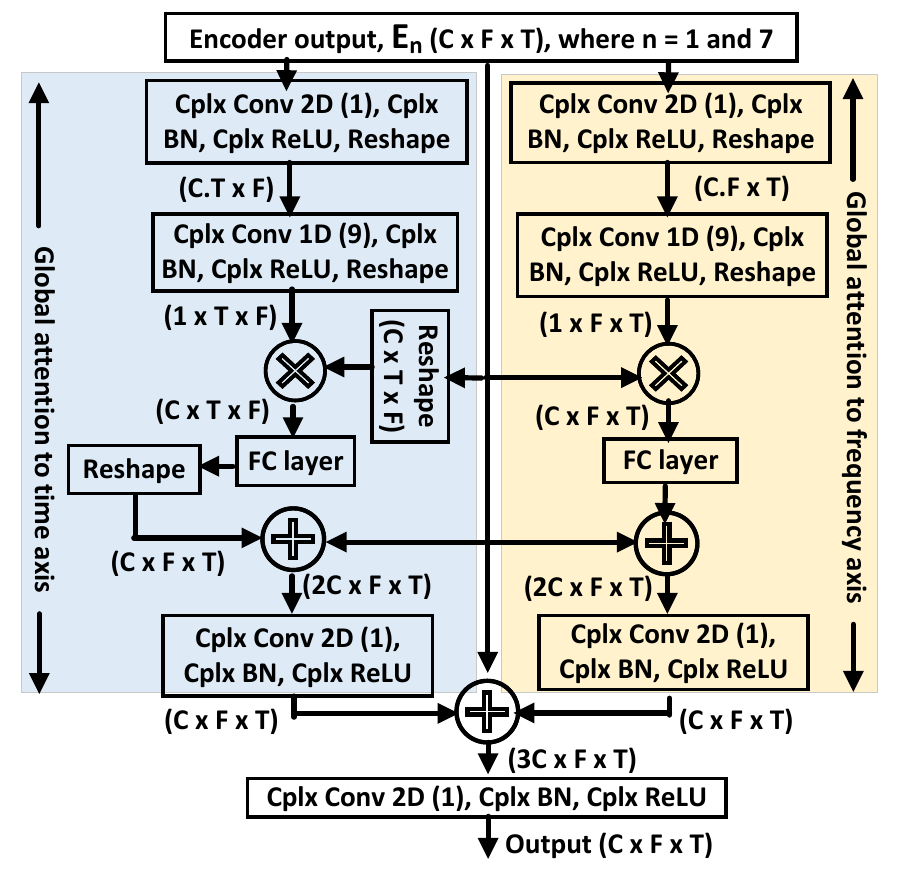} 
 \vspace{-01.05em}
  \caption{CGAB captures global time and frequency correlations. Here, Cplx = complex, and Conv = convolution.}
  \label{fig:CGAB}
  \vspace{-0.7515em}
\end{figure}

\textbf{Step 1 - Reshaping along the T-F axes:} The  encoder's output $E_n$ is decomposed into two tensors: one along the time axis and another along the frequency axis. Formally, $E_n$, which has a feature dimension of $C \times F \times T$, is given at the input of CGAB. At the first stage of reshaping,  $E_n$ is parallelly reshaped into $C.T$ vectors with dimension $C \cdot T \times F$ and into $C.F$ vectors with dimension $C \cdot F \times T$. This reshaping is done using 2D complex convolution, CBN, and complex ReLU activation, followed by vector reshaping. In the second stage of reshaping, $C \cdot T \times F$ is reshaped into $1 \times T \times F$ and  $C \cdot F \times T$ is reshaped into $1 \times F \times T$ using 1D complex convolution, CBN, complex ReLU activation, followed by vector reshaping. The tensors with dimension $1 \times F \times T$ capture the global harmonic correlation along the frequency axis, and $1 \times T \times F$ capture the global inter-phoneme correlation along the time axis. The captured features along the T-F axes and the original features from $E_n$ are point-wise multiplied together to generate a combined feature map with a dimension of $C \times T \times F$ and $C \times F \times T$ along T and F axes, respectively. This point-wise multiplication captures the inter-channel relationship between the encoder's output and complex T-F axes.

\textbf{Step 2 - Global attention along the T-F axes:} It is possible to treat the spectrogram as a 2D image and learn the correlations between every two pixels in the 2D image. However, this is computationally too costly and is not realistic. On the other hand, ideally, we can use self-attention \cite{ashish2017attention} to learn the attention map from two consecutive complex T-F spectrograms. But this might not be necessary. Because, on the time axis in each T-F spectrogram, when calculating SNR, the same set of parameters in the recursive relation is used, which suggests that temporal correlation is time-invariant among consecutive spectrograms. Moreover, harmonic correlations are independent in the consecutive spectrograms \cite{scalart1996speech}.

Based on this understanding, we propose a self-attention technique along the T-F axes. 
Specifically, attention to frequency and time axes is implemented by two separate fully connected (FC) layers. Along the time path, the input and output dimensions of FC layers are $C \times T \times F$. Along the frequency path, the input and output dimensions of FC layers are $C \times F \times T$. FC layer learns weights from complex T-F spectrograms and technically is different from the self-attention \cite{ashish2017attention} operation. To capture interchannel relationships among the input $E_n$ and output of FC layers, concatenation occurs, followed by 2D complex convolutions, CBN, and complex ReLU activation. Finally, the learned weights from the T-F axes are concatenated together to form a unified tensor, which holds joint information on the T-F global correlations from each spectrogram. 

We use only two CGABs - one after the 1st encoder, and another after the 7th encoder - to optimize the design. 

\vspace{-0.2em}
\subsection{Complex Multiresolution STFT Loss}
\label{subsec:Complex Multi-Resolution STFT Loss}
\vspace{-0.2em}


\mycolor{Unlike standard magnitude-only loss \cite{liu2022neural}, we propose a \emph{Complex Multi-Resolution STFT Loss} that separately evaluates the loss on real (i.e., amplitude) and imaginary parts (i.e., phase) of STFT across multiple resolutions, thereby capturing fine and coarse spectral details in complex domains.} At first, spectral convergence loss $L_{SC}$ \cite{tian2020tfgan} and log STFT magnitude loss $L_{mag}$ \cite{tian2020tfgan} are calculated on both real and imaginary parts of the enhanced signal and ground truth speech data. Let us define the $L_{SC}$ and $L_{mag}$ calculated on real and imaginary STFT data as \{$L^r_{SC}$, $L^i_{SC}$\} and \{$L^r_{mag}$, $L^i_{mag}$\}, respectively. Assuming we have $S$  different STFT resolutions, we aggregate the losses by averaging over the resolutions for both real and imaginary parts, following Eqn. \ref{eqn:multiresolutionstft}.

\vspace{-0.95em}
\begin{equation}
\small
\begin{aligned}
L_{\mathrm{r}} = \frac{1}{S} \sum_{s=1}^{S} \Big( L_{\mathrm{SC}}^{r} + L_{\mathrm{mag}}^{r} \Big); \,\,
  L_{\mathrm{i}} = \frac{1}{S} \sum_{s=1}^{S} \Big( L_{\mathrm{SC}}^{i} + L_{\mathrm{mag}}^{i} \Big)
\end{aligned}
\label{eqn:multiresolutionstft}
\vspace{-0.40em}
\end{equation}

We use $S$ = 3 resolutions, such as frequency bins = [256, 512, 1024], hop sizes = [128, 256, 512], and window lengths = [256, 512, 1024] to calculate $L_r$ and $L_i$. The overall complex multiresolution STFT loss, $L^{complex}_{STFT}$ is the sum of $L_r$ and $L_i$. The joint optimization in complex T-F domain on magnitude and phase improves the \textit{quality and intelligibility} of the speech reconstructed from pressure sensor data.

\vspace{-0.1em}
\section{Attack Model Demonstration in Cleanroom} 
\label{sec:pilot study}
\vspace{-0.1em}

We demonstrate our attack at an \textbf{FDA-compliant cleanroom located in an anonymous facility} shown in Fig.\ref{fig:experimentesetup} (Left). 
The facility uses an industry-used DPS from Sensiron with part\# SDP810-125PA \cite{datasheetsdp}. It has two input ports connected to two vinyl sampling tubes with inner diameters of 3/16" and 5/16" \cite{vinyltube}. A pressure pickup device with part\# A-417A \cite{pressurepickup} is connected to one input port (see Fig. \ref{fig:DPS_NPR}). A volunteer speaks from 0.5 m distance from the pressure pickup device. We record the output data from the DPS with a sampling frequency of 1 kHz. The recorded data from the DPS are shown in Fig. \ref{fig:experimentesetup} (Middle) with its matching ground-truth audio (see {\href{https://sites.google.com/view/pressuresensorwali/home}{\textcolor{blue}{{WaLi}}}). 

In Fig. \ref{fig:experimentesetup}, the time domain waveform indicates a strong correlation between the ground truth audio and the pressure sensor data. We measure the intelligibility of the pressure sensor data by taking the ground truth as a reference using a metric named Perceptual Evaluation of Speech Quality (PESQ). PESQ is widely used to evaluate how a degraded speech signal compares to a reference (ground truth) signal, closely modeling human perception. The PESQ for the pressure data is 0.98. PESQ can have values between -0.5 to 4.5, where -0.5 means low perceptual quality and 4.5 means perceptually close to the ground truth. The 0.98 of PESQ for the pressure data indicates that the DPS captures low-grade perceptual audio, which is not intelligible (to be intelligible PESQ value of more than 1.4 is required). Therefore, it indicates that DPS can be used to eavesdrop by an attacker if the attacker can reconstruct intelligible audio from the severely aliased and low-grade perceptual data collected from DPSs.


\begin{figure}[h]
\vspace{-0.920em}
\centering
\includegraphics[width=0.49\textwidth,height=0.11\textheight]{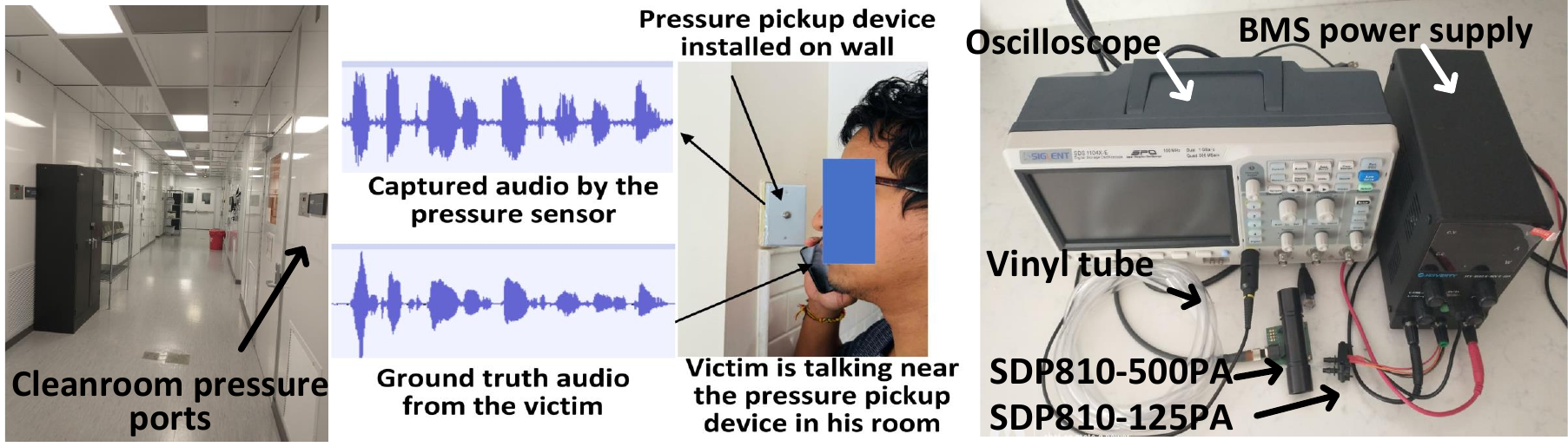}
\vspace{-01.8720em}
\caption{(Left, Middle) Demonstration of attack on real-world setup. (Right) Experimental setup for attack model evaluation using BMS and DPSs.}
\label{fig:experimentesetup}
\vspace{-0.8500em}
\end{figure}


\begin{table*}[ht!]
\vspace{-01.00em}
    \footnotesize
    \centering
    \setlength{\tabcolsep}{01.75pt}
     \caption{Evaluating WaLi for reconstructing speech from pressure sensor data for 500 Hz, 1 kHz, and 2 kHz to 8 kHz upsampling for 60 dB audio. Here, L = LSD, N = NISQA-MOS, S = SI-SDR, P = PESQ, and ST = STOI, \mycolor{W = Whisper, and A = AssemblyAI}.}
    \vspace{-01.00em}
    \begin{tabular}
     {c|c|c|c|c|c|c|c|c||c|c|c|c|c|c|c||c|c|c|c|c|c|c}
    \hline

          & \cellcolor [gray]{0.85}\textbf{Sensor} & \multicolumn{7}{c||} {\cellcolor [gray]{0.85}\textbf{2 kHz to 8 kHz}}  & \multicolumn{7}{|c||} {\cellcolor [gray]{0.85}\textbf{1 kHz to 8 kHz}} &  \multicolumn{7}{|c} {\cellcolor [gray]{0.85}\textbf{500 Hz to 8 kHz}} \\ 
         \hline
          &   & \textbf{L$\downarrow$}   & \textbf{N$\uparrow$} & \textbf{S$\uparrow$}  & \textbf{P$\uparrow$} & \textbf{ST$\uparrow$} & \multicolumn{2}{c||} {\textbf{\mycolor{WER (\%)}$\downarrow$}}  &  \textbf{L$\downarrow$}  & \textbf{N$\uparrow$} & \textbf{S$\uparrow$}  & \textbf{P$\uparrow$} & \textbf{ST$\uparrow$} & \multicolumn{2}{c||}{\textbf{\mycolor{WER (\%)}$\downarrow$}}  & \textbf{L$\downarrow$}   & \textbf{N$\uparrow$} & \textbf{S$\uparrow$}  & \textbf{P$\uparrow$} & \textbf{ST$\uparrow$} & \multicolumn{2}{c}{\textbf{\mycolor{WER (\%)}$\downarrow$}}\\
         
        \hline
        \multirow{3}{*} {Raw pressure data}  & 125PA & 2.95  &  1.27 & 9.87 & 1.14  &  0.79  & \mycolor{78(W)} & \mycolor{77(A)}  & 3.15 & 0.95 & 8.54  &  0.98 & 0.75 & \mycolor{85(W)} & \mycolor{86(A)} & 3.45 &  0.84 &  6.24  &  0.87 & 0.71 & \mycolor{97(W)} & \mycolor{98(A)}\\

         & \mycolor{Setra} & \mycolor{2.94}  &  \mycolor{1.29} & \mycolor{9.86} & \mycolor{1.13}  &  \mycolor{0.79}  & \mycolor{76(W)} & \mycolor{78(A)}  & \mycolor{3.14} & \mycolor{0.96} & \mycolor{8.57}  &  \mycolor{0.97} & \mycolor{0.75} & \mycolor{86(W)} & \mycolor{87(A)} & \mycolor{3.47} &  \mycolor{0.86} &  \mycolor{6.22}  &  \mycolor{0.88} & \mycolor{0.71} & \mycolor{98(W)} & \mycolor{99(A)}\\

          & \mycolor{500PA} & \mycolor{2.94}  &  \mycolor{1.29} & \mycolor{9.88} & \mycolor{1.18}  &  \mycolor{0.79}  & \mycolor{78(W)} & \mycolor{78(A)}  & \mycolor{3.16} & \mycolor{0.98} & \mycolor{8.57}  &  \mycolor{0.97} & \mycolor{0.75} & \mycolor{85(W)} & \mycolor{86(A)} & \mycolor{3.45} &  \mycolor{0.84} &  \mycolor{6.24}  &  \mycolor{0.87} & \mycolor{0.71} & \mycolor{98(W)} & \mycolor{97(A)}\\
        
        
        \hline

       \multirow{2}{*} {Reconstructed} & 125PA & 1.01  &  2.17 & 11.38 & 1.95  &  0.82  & \mycolor{20(W)} & \mycolor{21(A)} & 1.16 & 1.95   &  10.04    &  1.61    &  0.79 & \mycolor{26(W)}  & \mycolor{27(A)} & 1.24 &  1.78  & 8.78  &  1.51    & 0.75  & \mycolor{38(W)} & \mycolor{39(A)}\\ 
       
        & \mycolor{Setra} & \mycolor{1.00}  &  \mycolor{2.18} & \mycolor{11.37} & \mycolor{1.96}  &  \mycolor{0.82}  & \mycolor{21(W)} & \mycolor{20(A)} & \mycolor{1.15} & \mycolor{1.96}   &  \mycolor{10.02}    &  \mycolor{1.63}    &  \mycolor{0.79} &  \mycolor{27(W)} & \mycolor{27(A)} & \mycolor{1.26} &  \mycolor{1.80}  & \mycolor{8.79} &  \mycolor{1.53}    & \mycolor{0.75}  & \mycolor{38(W)} & \mycolor{38(A)}\\
        
        & \mycolor{500PA} & \mycolor{1.00}  &  \mycolor{2.16} & \mycolor{11.37} & \mycolor{1.97}  &  \mycolor{0.82}  & \mycolor{21(W)}  & \mycolor{21(A)} & \mycolor{1.17} & \mycolor{1.94}   &  \mycolor{10.14}    &  \mycolor{1.60}    &  \mycolor{0.79} & \mycolor{26(W)}  & \mycolor{27(A)} & \mycolor{1.24} &  \mycolor{1.80}  & \mycolor{8.79}  &  \mycolor{1.53}    & \mycolor{0.75}  & \mycolor{36(W)} & \mycolor{38(A)}\\



    \end{tabular}
    \label{table:BWEwithdiff_freq}
    \vspace{-02.200em}
\end{table*}

\vspace{-0.31em}
\section{Attack Model Evaluation}
\label{sec:Attack Model Evaluation}
\vspace{-0.1em}

We demonstrate WaLi at an FDA-compliant anonymous cleanroom in Section \ref{sec:pilot study}. As it was not \textit{permitted} to experiment with the HVAC located in the cleanroom for collecting large corpus of pressure data, we prepare a testbed to evaluate WaLi using the same DPS (part\# SDP810-125PA), vinyl tubes, and pressure pickup device  (see Section \ref{sec:pilot study}), shown in Fig. \ref{fig:experimentesetup} (Right). \mycolor{We also verify our work using two more industry-used DPSs -- SETRA264 \cite{setra264} from Setra Systems (Fig. \ref{fig:DPS_NPR} (Right)) and SDP810-500PA \cite{datasheetsdp} (see Appendix \ref{appensec:5SensorDetails} for sensor details).} 

\vspace{-0.20em}
\subsection{Pressure Data and Audio Corpus} 
\label{subsec:Speech_Corpus}
\vspace{-0.1em}

We collect pressure data in two different ways. \textbf{First}, we use 20 volunteers (9 male and 11 female) to utter from  Wikipedia and collect a \mycolor{total of 8.33 hours of pressure} data from each of the three DPSs 
($\sim$25 minutes from each volunteer with permission and no ethical concern). \textbf{Second}, as \mycolor{8.33 hours} of data is not enough for proper model training, we use VCTK v0.92 \cite{yamagishi2019cstr}, a multi-speaker English corpus, and play the audio at 60 dB through a loudspeaker and collect the corresponding pressure data for evaluation. The dataset contains $\sim$39 hours of 
audio traces sampled at 48 kHz from 110 individuals (i.e., balanced for males and females, open source and no ethical concern). We downsample the dataset to 8 kHz for evaluation. Each audio clip has a duration ranging from 2s to 7s. We standardize all audio clips to 4s by either zero-padding or silence trimming. 
The speaker is placed at a 0.5 m distance from one of the pressure ports to play the audio clips.  To prevent sound signals from influencing the pressure in the other port, we connect a sampling tube to it, isolated by positioning its opening a meter away. Signals from the sensor are collected as .edf and next converted to .wav format using a custom Python script. The sampling frequency of the DPS is varied in between 500 Hz and 2 kHz. We use sinc interpolation to upsample before the audio reconstruction to ensure that the system input and output have the same shape.  \mycolor{Note that in a real case, the speech contents may be different from the spoken ones during the attack phase. Thus, for testing, we use 11 different speakers not present in the training to prove that WaLi reconstructs speech from the victim's pressure sensor data without using the ground truth audio of the victim.}  The models are trained offline with an NVIDIA 4090 GPU. 

\vspace{-0.4em}
\subsection{Comprehensive Evaluation Metrics}
\label{subsec:Evaluation_Metrics}
\vspace{-0.31em}

\mycolor{To comprehensively evaluate the reconstructed speech, we use six evaluation metrics: log spectral distance (LSD) \cite{liu2022neural} for spectral distortion, short-time objective intelligibility (STOI) \cite{taal2011algorithm} for intelligibility, perceptual evaluation of speech quality (PESQ) \cite{rix2001perceptual} for perceived quality, scale-invariant signal-to-distortion ratio (SI-SDR) \cite{le2019sdr} for overall signal distortion, non-intrusive speech quality assessment - mean opinion score (NISQA-MOS) and word error rate (WER) for perceived quality. We refer to Appendix \ref{append:Evaluation_Metrics} for more details on metrics.}

\vspace{-0.3em}
\subsection{Overall Performance}
\label{sec:Overall Performance}
\vspace{-0.1em}

To \textit{intuitively} observe the performance of WaLi, we compare among ground truth speech, raw data from the pressure sensor, and reconstructed speech by WaLi (see Fig. \ref{fig:comparison}). The frequency components above 250 Hz are absent in the pressure sensor data (see Fig. \ref{fig:comparison}(b)) as we use a sampling frequency of 0.5 kHz in DPS at this time. The ground truth audio is sampled at 8 kHz (4 kHz bandwidth), shown in Fig. \ref{fig:comparison}(a). From Fig. \ref{fig:comparison}(c), we can see that the high-frequency components are reconstructed up to 4 kHz by WaLi. The reconstructed spectrogram shows a high similarity to the ground truth one.

Moreover, to \textit{quantitatively} measure WaLi's performance,  we vary the sampling frequency of the pressure sensor on three scales -- 500 Hz, 1 kHz, and 2 kHz. We reconstruct the audio to 8 kHz, to evaluate WaLi for high upsampling ratios \cite{liu2022neural}. Table \ref{table:BWEwithdiff_freq} shows the results \mycolor{for three industry-used DPSs -- SDP810-125PA, SDP810-500PA, and SETRA264 -- for \textit{combined} 8.33-hour HVAC data and VCTK dataset. Table \ref{table:BWEwithdiff_freq} indicates that WaLi's performance does not vary for different sensors.} WaLi achieves a 2.78x improvement in LSD (i.e., 3.45 vs 1.24), 2.11x increase in NISQA-MOS (i.e., 0.84 vs 1.78), 1.4x increase in SI-SDR (i.e., 6.24 vs 8.78), 1.73x increase in PESQ (i.e., 0.87 vs 1.51), and 1.05x increase in STOI (i.e., 0.71 vs 0.75) between the raw pressure data and the reconstructed audio for 500 Hz sampling \mycolor{for SDP810-125PA.}

Please also note that the metrics are more improved for 2-8 kHz compared to 1-8 kHz and 0.5-8 kHz because for 2-8 kHz, the upsampling ratio is around 4, for 1-8 kHz the upsampling ratio is 8, and for 0.5-8 kHz the upsampling ratio is 16. The lower the upsampling ratio, the greater the performance gain for the reconstruction model.

\vspace{-0.8705em}
\begin{figure}[ht!]
  \centering
 \includegraphics[width=0.48\textwidth,height=0.11\textheight]{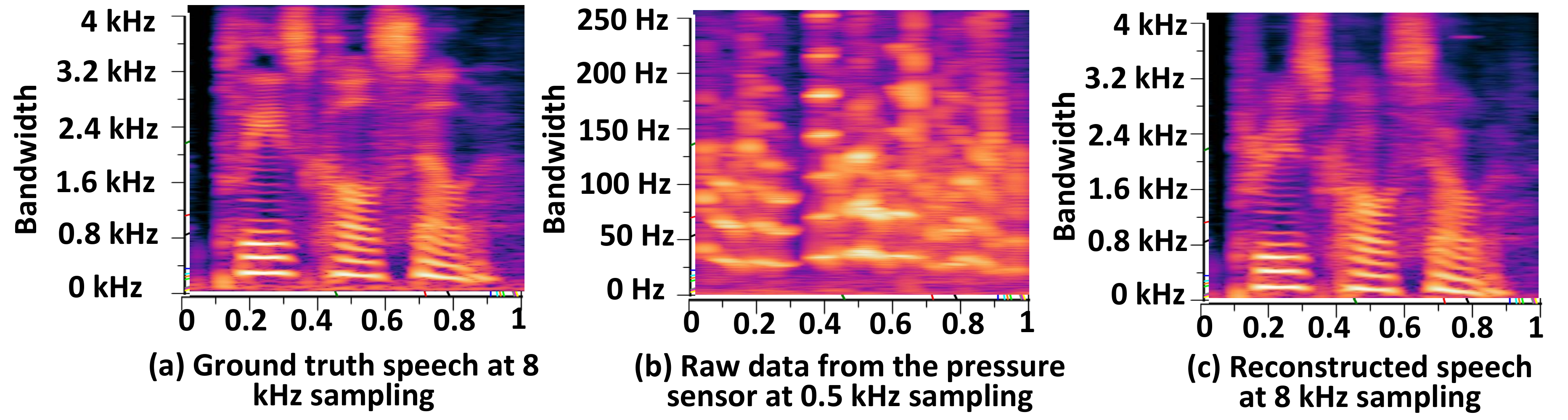} 
 \vspace{-01.205em}
  \caption{The reconstructed spectrogram from the pressure sensor data in (c) shows a high similarity to the ground truth one in (a).}
  \label{fig:comparison}
  \vspace{-01.0em}
\end{figure}

\mycolor{
\textbf{Testing on real-world HVAC dataset:} Table \ref{table:BWEwithdiff_freq} shows results for the \textit{combined} \{8.33-hours real-world HVAC and the VCTK\} test dataset. We also evaluate WaLi \textit{separately} on 20\% of 8.33-hour HVAC data and 20\% of the VCTK dataset, shown in Table \ref{table:seperatedataset}. We see that WaLi gives similar performance for both the real-world HVAC and the VCTK dataset, indicating WaLi's effectiveness on real-world HVAC systems.}

\begin{table}[ht!]
\vspace{-01.200em}
    \footnotesize
    \centering
    \setlength{\tabcolsep}{04.10pt}
     \caption{\mycolor{Evaluating separately on R = Real-world HVAC data, and V = VCTK dataset for 500 Hz to 8 kHz upsampling.}} 
    \vspace{-01.00em}
    {\color{red}
    \begin{tabular}
     {c|c|c|c|c|c|c|c|c|c|c}
    \hline

          \cellcolor [gray]{0.85}\textbf{Phase} & \multicolumn{2}{c|} {\cellcolor [gray]{0.85} \textbf{L$\downarrow$}}  & \multicolumn{2}{c|} {\cellcolor [gray]{0.85} \textbf{N$\uparrow$}} & \multicolumn{2}{c|} {\cellcolor [gray]{0.85} \textbf{S$\uparrow$}}  & \multicolumn{2}{c|} {\cellcolor [gray]{0.85} \textbf{P$\uparrow$}} & \multicolumn{2}{c} {\cellcolor [gray]{0.85} \textbf{ST$\uparrow$}}\\
          \hline
             & \cellcolor [gray]{0.85} \textbf{R}   & \cellcolor [gray]{0.85} \textbf{V} & \cellcolor [gray]{0.85} \textbf{R} & \cellcolor [gray]{0.85} \textbf{V}  & \cellcolor [gray]{0.85} \textbf{R} & \cellcolor [gray]{0.85} \textbf{V} & \cellcolor [gray]{0.85} \textbf{R} & \cellcolor [gray]{0.85} \textbf{V} & \cellcolor [gray]{0.85} \textbf{R} & \cellcolor [gray]{0.85} \textbf{V} \\
         
        \hline

        WaLi & 1.24 & 1.23 &  1.78 & 1.78 & 8.78 & 8.77 & 1.50 & 1.51 &  0.7 &  0.75\\ 
    \end{tabular}
    }
    \label{table:seperatedataset}
    \vspace{-01.00em}
\end{table}

\textbf{Individual test speaker:} In addition to average values, Fig. \ref{fig:comparisonindividual} (Left) shows the individual values of each test speaker for LSD and NISQA-MOS. The individual values indicate that WaLi performs consistently for all unseen test speakers. 

\mycolor{\textbf{Subjective analysis:} To assess the perceptual quality of WaLi relative to the raw pressure data, we conduct a listening test with 18 participants using a 5-point Mean Opinion Score (MOS) scale (1 = poor, 5 = excellent). As shown in Fig. \ref{fig:comparisonindividual} (Right), the average MOS of $\sim$2.7 for 500 Hz to 8 kHz and  3.8 for 2 kHz to 8 kHz upsampling indicates that the reconstructed speech achieves a \textit{fair to moderate quality} relative to the \textit{completely unrecognizable} raw pressure data (i.e., MOS = $\sim$1). This finding is also supported by the WER discussed next. We refer to Appendix \ref{appensubsec:subj_eval_details} for the experimental setup for MOS test.} 

\mycolor{\textbf{Speech-to-Text accuracy:} We evaluate the reconstructed speech using two pretrained models—OpenAI’s Whisper-base \cite{whisper} and the AssemblyAI's Speech-to-Text Recognition API \cite{AssemblyAi}—to compute the WER. 
The resulting WERs (see Table \ref{table:BWEwithdiff_freq})  are approximately $\sim$20\%, $\sim$26\%, and $\sim$38\% for sampling frequencies of 2 kHz, 1 kHz, and 500 Hz, respectively, substantially lower than the $\sim$78-98\% WER observed for the raw pressure data.  These findings reinforce that WaLi consistently reconstructs higher perceptual quality speech favored by human listeners,  transforming pressure signals that were otherwise \textit{completely incomprehensible} in their raw form.}



\begin{figure}[ht!]
 \vspace{-01.315em}
  \centering
 \includegraphics[width=0.49\textwidth,height=0.12\textheight]{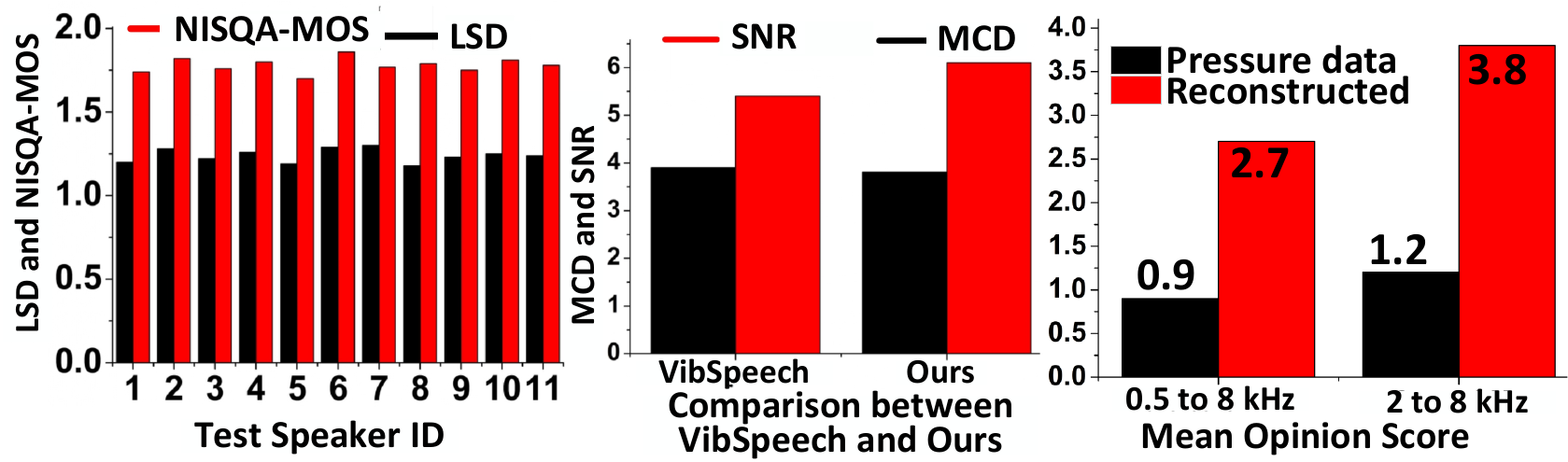} 
 \vspace{-01.71em}
  \caption{(Left) LSD and NISQA-MOS for individual test speakers. (Middle) Comparison between WaLi and VibSpeech \cite{wang2024vibspeech}. \mycolor{(Right) MOS score.}}
  \label{fig:comparisonindividual}
  \vspace{-0.515em}
\end{figure}

\textbf{Comparison with VibSpeech \cite{wang2024vibspeech}:} We also compare between WaLi and VibSpeech \cite{wang2024vibspeech} in terms of SNR and Mel-Ceptral Distortion (MCD) \cite{wang2024vibspeech} in Fig. \ref{fig:comparisonindividual} (Middle), which indicates that WaLi performs better (MCD = 3.61 and SNR = 6.3 dB) than VibSpeech (MCD = 3.7 and SNR = 6.1 dB).

Please note that WaLi is tested with different speakers, which are not seen by the model during the training time. Moreover, the VCTK  has everyday-use audio clips, which are not limited to a certain specialized vocabulary, such as \cite{achamyeleh2024fly}. \mycolor{Therefore, WaLi can recover moderately intelligible audio with unrestricted vocabulary, irrespective of applications.}

\begin{table}[ht!]
\vspace{-01.500em}
    \footnotesize
    \centering
     \caption{Evaluation of phase reconstruction at non-noisy conditions.} 
    \vspace{-01.00em}
    \begin{tabular}
     {c|c|c|c|c|c}
    \hline
          \cellcolor [gray]{0.85}\textbf{Phase} & \cellcolor [gray]{0.85} \textbf{L$\downarrow$}   & \cellcolor [gray]{0.85} \textbf{N$\uparrow$} & \cellcolor [gray]{0.85} \textbf{S$\uparrow$}  & \cellcolor [gray]{0.85} \textbf{P$\uparrow$} & \cellcolor [gray]{0.85} \textbf{ST$\uparrow$}\\
        \hline
         Griffin-Lim &  1.26  &  1.71 & 7.65 & 1.47  &  0.74 \\
        \hline
        WaLi & 1.24  &  1.78 & 8.78 & 1.51  &  0.75   \\
    \end{tabular}
    \label{table:phase-nonnoiy}
    \vspace{-01.200em}
\end{table}

\vspace{-0.3em}
\subsection{Phase Reconstruction in Non-Noisy Condition}
\label{sec:Impact of Phase Reconstruction}
\vspace{-0.1em}

Previous works used the Griffin-Lim algorithm \cite{hu2022accear} as a vocoder to reconstruct the phase. To compare the performance of our phase reconstruction in the non-noisy condition, we reconstruct audio from only the magnitude of our model with the Griffin-Lim algorithm, and then compare it with the reconstructed audio from our model. A comparison summary is shown in Table \ref{table:phase-nonnoiy} for a 500 Hz to 8 kHz reconstruction. Table \ref{table:phase-nonnoiy} indicates that our proposed complex-valued WaLi is similar/slightly better in phase reconstruction compared to the Griffin-Lim algorithm in the non-noisy condition. However,  the performance of WaLi is much better under noisy conditions that is discussed in the next section.

\begin{table}[ht!]
\vspace{-01.200em}
    \footnotesize
    \centering
    \setlength{\tabcolsep}{03.0pt}
     \caption{\mycolor{Evaluation of phase reconstruction at noisy conditions separately for R = Real-world HVAC data \& D = DCASE data.}} 
    \vspace{-01.00em}
    {\color{red}
    \begin{tabular}
     {c|c|c|c|c|c|c|c|c|c|c}
    \hline

          \cellcolor [gray]{0.85}\textbf{Phase} & \multicolumn{2}{c|} {\cellcolor [gray]{0.85} \textbf{L$\downarrow$}}  & \multicolumn{2}{c|} {\cellcolor [gray]{0.85} \textbf{N$\uparrow$}} & \multicolumn{2}{c|} {\cellcolor [gray]{0.85} \textbf{S$\uparrow$}}  & \multicolumn{2}{c|} {\cellcolor [gray]{0.85} \textbf{P$\uparrow$}} & \multicolumn{2}{c} {\cellcolor [gray]{0.85} \textbf{ST$\uparrow$}}\\
          \hline
             & \cellcolor [gray]{0.85} \textbf{R}   & \cellcolor [gray]{0.85} \textbf{D} & \cellcolor [gray]{0.85} \textbf{R} & \cellcolor [gray]{0.85} \textbf{D}  & \cellcolor [gray]{0.85} \textbf{R} & \cellcolor [gray]{0.85} \textbf{D} & \cellcolor [gray]{0.85} \textbf{R} & \cellcolor [gray]{0.85} \textbf{D} & \cellcolor [gray]{0.85} \textbf{R} & \cellcolor [gray]{0.85} \textbf{D} \\
         
        \hline
         Griffin-Lim &  1.38 & 1.40 &  1.43 & 1.41 & 6.17 & 6.19 & 1.21  & 1.23 &  0.70 & 0.69\\
        \hline

        WaLi & 1.26 & 1.28 &  1.71 & 1.73 & 8.14 & 8.18 & 1.47 &  1.49 &  0.74 & 0.73 \\ 
    \end{tabular}
    }
    \label{table:phase-noisy}
    \vspace{-01.200em}
\end{table}

\vspace{-0.85em}
\subsection{Phase Reconstruction in Noisy Conditions}
\label{sec:Impact of Phase Reconstruction noisy}
\vspace{-0.2em}

DPSs, operating within a low-pressure range of 0-200 Pa and at high sampling frequencies of 0.5-2 kHz, are sensitive to transient noise sources (see Section \ref{subsecc:Phase Reconstruction for Intelligible Speech}). Therefore, clean phase reconstruction becomes particularly significant in noisy conditions. \mycolor{To test the performance of phase reconstruction under noisy conditions, we use 80\% of the \textbf{8.33-hour} dataset collected previously from the real-world HVACs to train the model. This dataset has the actual noisy condition of a real-world HVAC system. Moreover, as this real-world HVAC noisy data is not enough to train our model, we also use the DCASE challenge dataset \cite{mesaros2016tut} to add transient mechanical noise, such as fans, valves, pump noise, air leaks, tool drops, etc., to the clean VCTK dataset. Noise is randomly added within a range of -15 to 15 dB. Then, we retrain WaLi with a \{noisy, clean\} pair combined from both the real-world and DCASE datasets. We test WaLi \textit{separately} using the real-world noisy data collected from the HVAC and DCASE dataset.} A comparison summary is shown in Table \ref{table:phase-noisy} for a 500 Hz to 8 kHz reconstruction in noisy conditions. Table \ref{table:phase-noisy} indicates that our proposed complex-valued WaLi is much better in phase reconstruction compared to the Griffin-Lim algorithm in transient noisy conditions \mycolor{for both real-world HVAC and DCASE datasets}. The reason behind this is that WaLi handles both magnitude and phase jointly and recovers both clean magnitude and phase under noisy conditions from the clean reference signal using a complex multi-resolution loss function. On the other hand, Griffin-Lim-based models \cite{hu2022accear} do not learn clean phase reconstruction as these models are magnitude-only models and cannot handle phase jointly.

  \vspace{-01.39515em}
\begin{figure}[ht!]
  \centering
\includegraphics[width=0.44\textwidth,height=0.13\textheight]{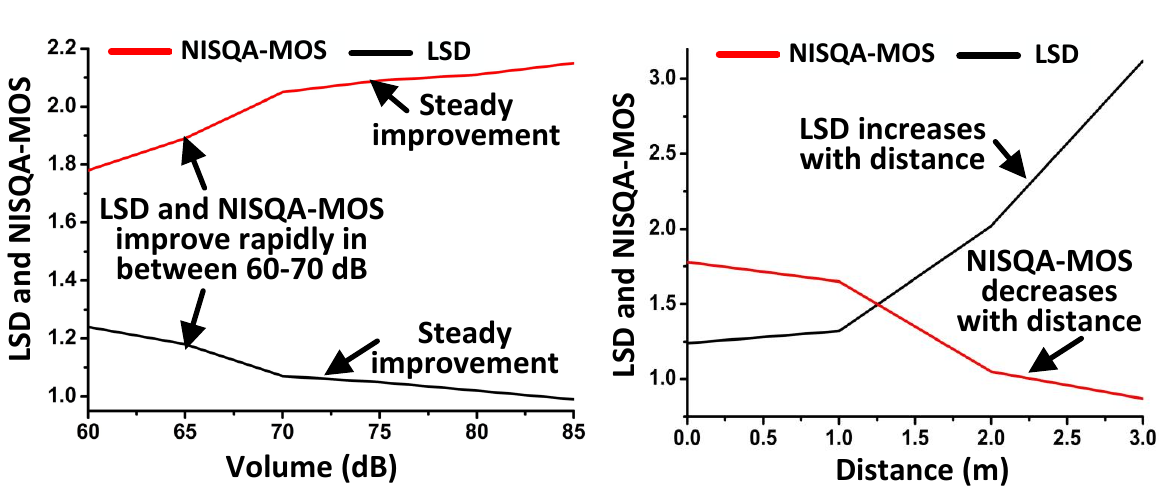} 
 \vspace{-01.15em}
  \caption{(Left) Relationship with speaker volume and (Right) speaker distance with the reconstruction performance.}
  \label{fig:impactofvolume}
  \vspace{-0.915em}
\end{figure}

\vspace{-0.2em}
\subsection{Impact of Sound Pressure Level}
\label{subsec:Impact of Sound Volume}
\vspace{-0.2em}

We  vary the SPL from 60 dB to 85 dB with a 5 dB increment. We inject the audio into the pressure sensor ports through the 1m long sampling tube from a 0.5 m distance. The result is shown in Fig. \ref{fig:impactofvolume} (Left) for LSD and NISQA-MOS for 500 Hz to 8 kHz upsampling. We can see that the LSD and NISQA-MOS improve rapidly between 60 dB and 70 dB. 
\mycolor{However, after 70 dB, the increase in volume does not drastically change the low-resolution frequency components. Therefore, the improvement is not sharp after 70 dB; rather, a steady ascent occurs.}  
The LSD is close to 1, and NISQA-MOS is close to 2 for a wide volume range from normal (i.e., 60 dB) to loud speech, indicating WaLi's capability to reconstruct intelligible speech from normal to loud in real-world scenario.

\vspace{-0.2em}
\subsection{Impact of Speaker Distance}
\label{subsec:Impact of Speaker Distance}
\vspace{-0.2em}

We vary the distance of a speaker from 0 m (touching the pressure sensor's input port) to 3 m from the target pressure sensor. The result is shown in Fig. \ref{fig:impactofvolume} (Right) for LSD and NISQA-MOS for 500 Hz to 8 kHz upsampling for 60 dB audio. 
\mycolor{The improvement in LSD and NISQA-MOS decreases with increasing distance from the audio source because of the inverse-proportional law \cite{soundpropagation}.} 
It is also clear that WaLi performs well up to 1.2 m distance. After 1.2 m, the reconstructed audio has severely degraded intelligibility.

  \vspace{-01.15em}
\begin{figure}[ht!]
  \centering
 \includegraphics[width=0.49\textwidth,height=0.1\textheight]{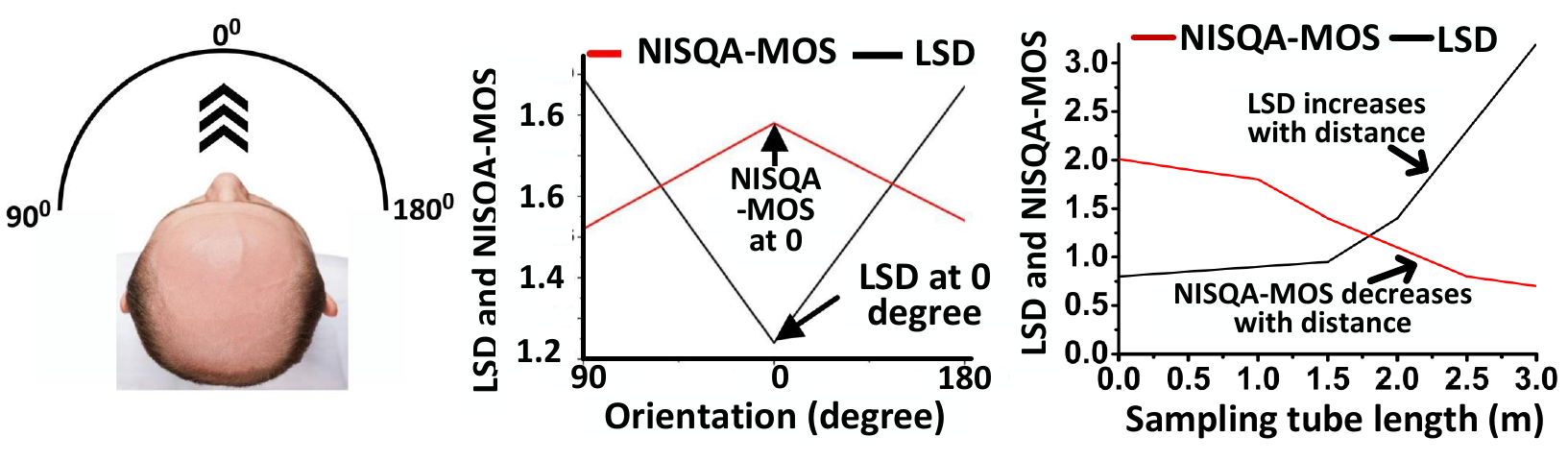} 
 \vspace{-02.515em}
  \caption{(Left, Middle) Impact of speaker orientation on the audio reconstruction quality. \mycolor{(Right) Impact of sampling tube length.}}
  \label{fig:impactoforientation}
  \vspace{-0.915em}
\end{figure}

\vspace{-0.2em}
\subsection{\mycolor{Impact of Sampling Tube Length}}
\label{subsec:Impact of Sampling Tube Length}
\vspace{-0.2em}

\mycolor{
We vary the sampling tube length from 0 m (touching the pressure port) to 3 m with a 0.5 m increment for 60 dB audio and 500 Hz to 8 kHz upsampling. Fig. \ref{fig:impactoforientation} (Right) shows that LSD and NISQA-MOS degrade with the increase of the tube length as sound energy attenuates with the increase of the tube length, similar to speaker distance evaluation in Section \ref{subsec:Impact of Speaker Distance}.}

\vspace{-0.2em}
\subsection{Impact of Speaker Orientation}
\label{subsec:Impact of Speaker's Orientation}
\vspace{-0.2em}

We place the speaker at 0\textdegree (in front),  90\textdegree (left), and 180\textdegree  (right) of the pressure sensor at 0.5 m away and generates a sound of 60 dB. The results are shown in Fig. \ref{fig:impactoforientation} (Middle) for LSD and NISQA-MOS for 500 Hz to 8 kHz upsampling. The LSD and NISQA-MOS are higher at 0\textdegree \,orientation, as at  0\textdegree,  the audio is directed to the pressure sensor ports with minimum loss, potentially creating a strong vibration.  Performance is the same at both 90\textdegree    and 180\textdegree orientations, but lower compared to 0\textdegree with still intelligible quality (NISQA-MOS = 1.5).

\begin{table}[ht!]
\vspace{-01.200em}
    \footnotesize
    \centering
    \renewcommand{\arraystretch}{0.85}
    \setlength{\tabcolsep}{02.20pt}
     \caption{\mycolor{Multi-speaker environment for 500 Hz to 8 kHz upsampling.}} 
    \vspace{-01.00em}
    {\color{red}
    \begin{tabular}
     {c|c|c|c|c|c|c}
    \hline

          \cellcolor [gray]{0.85}\textbf{Multi-speaker} & \multicolumn{3}{c|} {\cellcolor [gray]{0.85} \textbf{LSD$\downarrow$}}  & \multicolumn{3}{c} {\cellcolor [gray]{0.85} \textbf{NISQA-MOS$\uparrow$}} \\ 
          \hline
             & \cellcolor [gray]{0.85} \textbf{$\sim$15dB}   & \cellcolor [gray]{0.85} \textbf{$\sim$40dB} & \cellcolor [gray]{0.85} \textbf{$\sim$50dB} & \cellcolor [gray]{0.85} \textbf{$\sim$15dB}  & \cellcolor [gray]{0.85} \textbf{$\sim$40dB} & \cellcolor [gray]{0.85} \textbf{$\sim$50dB} \\ 
         
        \hline
         2 speakers &  1.21 & 1.37 & 1.65  & 1.72 & 1.61 & 1.31 \\
        \hline
        3 speakers &  1.38 & 1.52 &  1.81 & 1.65 & 1.47 & 1.14 \\
        \hline
        4 speakers &  1.59 & 1.67 &  1.93 & 1.41 & 1.29 &  0.92\\
    \end{tabular}
    }
    \label{table:multispeaker}
    \vspace{-01.200em}
\end{table}

\vspace{-0.52em}
\subsection{\mycolor{Impact of Multi-Speaker Environment}}
\label{subsec:Impact of Multi-Speaker Environment}
\vspace{-0.2em}

\mycolor{
We separately collect another 300 minutes of pressure data from
20 volunteers (9 male and 11 female) with two (one target speaker and one interfering speaker), three (one target speaker and two interfering speakers), and four (one target speaker and three interfering speakers) multi-speaker setups. The target speaker has an SPL of $\sim$60 dB, and the interfering speakers have SPLs of $\sim$15, $\sim$40, and $\sim$50 dB at 0.5m distance. We retrain WaLi using these multi-speaker data pairs (see Table \ref{table:multispeaker} for results). We see that with the increase in interference speakers and their SPLs, the reconstruction performance degrades. The reason is that when interfering speakers have SPLs close to the target speaker, the problem transforms from speech reconstruction to a speaker separation problem. WaLi performs well for 15 dB multi-speaker setup but its performance degrades for $>$15 dB multi-speaker setup as WaLi is not optimized for speaker separation tasks. One alternative approach is that we can use speaker separation algorithms \cite{nachmani2020voice} first and then use WaLi for speech reconstruction. The performance of this approach will largely depend on how well the speaker separation algorithm performs because we already show that WaLi works well for speech reconstruction.}

\vspace{-0.2em}
\subsection{Fine-Tuning on Victim's Ground Truth Audio}
\label{subsec:Impact of Pretraining to Speaker's Ground Truth}
\vspace{-0.2em}

Please note that  WaLi can reconstruct the audio even if the speaker is different as WaLi is tested with different speakers, which are not seen by the model during the training time.  This makes our attack model more flexible compared to prior works \cite{wang2024vibspeech, hu2022accear}, \mycolor{where prior works assume that clean ground-truth speech of the victim is available via a microphone.} 
However, if the attacker could manage the victim's \{ground truth, pressure sensor data\} pair, the accuracy of the models would be greatly improved by WaLi. To evaluate this,  we incorporate a small amount of  \{ground truth, pressure sensor data\} pairs from the victim to fine-tune the pre-trained network. As shown in Table \ref{table:finrtune} for 500 Hz to 8 kHz upsampling, the performance gradually improves if the amount of data from the victim increases. For example, with just one minute of data, we observe significant performance gains, and after five minutes of additional data, the performance gain is minimal. This proves that WaLi can be fine-tuned with the victim's audio, if possible, under a stronger assumption of the attack model. 

\begin{table}[ht!]
\vspace{-01.1800em}
    \footnotesize
    \centering
    \renewcommand{\arraystretch}{0.85}
        \caption{Evaluation after fine-tuning on victim's ground truth audio.} 
    \vspace{-0.900em}
    \begin{tabular}
     {m{3cm}|c|c|c|c|c}
    \hline

             \cellcolor [gray]{0.85}\textbf{Fine tuning data size} &  \cellcolor [gray]{0.85} \textbf{L$\downarrow$}   & \cellcolor [gray]{0.85} \textbf{N$\uparrow$} & \cellcolor [gray]{0.85} \textbf{S$\uparrow$}  & \cellcolor [gray]{0.85} \textbf{P$\uparrow$} &  \cellcolor [gray]{0.85} \textbf{ST$\uparrow$} \\
         
        \hline
        \hline
         0 min (before fine-tuning)  &  1.24   & 1.78  & 8.78 & 1.51 & 0.75  \\
        \hline
         1 min   &  1.03   & 1.89  & 10.23 & 1.72 & 0.77  \\             
\hline
 5 min   &  0.98   & 2.04  & 11.45 & 1.91 & 0.79  \\ 
           \hline
            7 min   &  0.97   & 2.07  & 11.53 & 1.92 & 0.79  \\ 

        \hline
    \end{tabular}
    
    \label{table:finrtune}
    \vspace{-0.95800em}
\end{table}

\vspace{-0.5em}
\subsection{Testing With Less Than 500 Hz Sampling}
\label{subsec:Impact500hz}
\vspace{-0.2em}

We test WaLi while reconstructing speech from a 250 Hz sampling frequency of the pressure sensor data, shown in Fig. \ref{fig:250hz} in Appendix \ref{appensec:500hzsampling}. As only a few pitch frequencies are present within 125 Hz bandwidth of the 250-Hz-sampled data, the reconstructed speech in Fig. \ref{fig:250hz}(c) is not intelligible with an NISQA-MOS value of 0.48. 
However, the 250-Hz or less than 250-Hz-sampled data may be used for hot word detection \cite{yao2024watch}. 

 \vspace{-01.20em}
\begin{table}[ht!] 
    \footnotesize
    \centering
    \setlength{\tabcolsep}{03.90pt}
    \renewcommand{\arraystretch}{0.85}
    \caption{\mycolor{Ablation study for 500 Hz to 8 kHz reconstruction.}}
    \vspace{-0.90em}
    {\color{red}
    \begin{tabular}{c|m{13em}|c|c|c|c|c}
        \hline
           \textbf{Row} & \textbf{Model}   & \textbf{L $\downarrow$} & \textbf{ST $\uparrow$} & \textbf{P $\uparrow$} & \textbf{S $\uparrow$} & \textbf{N $\uparrow$}  \\
        \hline
        \hline
         (i) & Real-valued net & 1.58 & 0.71 & 1.35 & 4.24 & 1.43  \\
        \hline
         (ii) & Remove CGAB   & 1.32 & 0.74 & 1.45 & 7.54 & 1.48 \\
         \hline 
        \multirow{2}{*} {(iii)}  & Transformer in bottleneck  & 1.34 &  0.75 & 1.41 & 7.77 & 1.65 \\
          & CNN in bottleneck & 1.43 & 0.73 & 1.38 & 7.94 & 1.69 \\
        \hline
         (iv) & Magnitude-only loss & 1.33 & 0.73 & 1.32 & 7.44 & 1.59 \\
        \hline
        \rowcolor{black!30}
         & \textbf{WaLi (complex-valued + conformer+CGAB+complex loss)}  &  \textbf{1.24} &  \textbf{0.75} &  \textbf{1.52} &  \textbf{8.78} &  \textbf{1.78} \\
        \hline
    \end{tabular}%
   }
    \label{tab:ablation}
    \vspace{-01.0em}
\end{table}

\vspace{-0.5em}
\subsection{\mycolor{Ablation study}}
\label{subsec:Ablationstudy}
\vspace{-0.2em}

\mycolor{
To understand the different parts of WaLi, we do an ablation study (see Table \ref{tab:ablation}) in the following four avenues: (i) We compare complex-valued WaLi with the real-valued WaLi in noisy conditions. We see that complex-valued WaLi does better in phase reconstruction and outperforms real-valued networks in real-world noisy conditions (see Sections \ref{sec:Impact of Phase Reconstruction} and \ref{sec:Impact of Phase Reconstruction noisy} for details). (ii) We remove CGAB from WaLi and observe that it does not perform well when CGAB is removed because of the lacking of capturing long-range correlations in time and frequency axes (see Section \ref{subsec:CGAB} for details). (iii) We separately use a transformer and CNN at the bottleneck layer and observe that they do not perform well compared to conformer (refer to Section \ref{subsec:conformer_symbolic} for the reasons). And (iv) We use magnitude-only losses and find that it does not perform well because magnitude-only losses can not reconstruct clean phases directly in noisy conditions.
}

\vspace{-0.5em}
\section{Limitation and Discussion}
\label{sec:Limitation and Discussion}
\vspace{-0.2em}


\mycolor{
\textbf{Feasibility of the Attack:} For a successful attack, the attack constraints are: distance $\leq$1.2 m, orientation near $0^0$, volume $\geq$60 dB, and sampling $\geq$500 Hz (see Sections \ref{subsec:Impact of Sound Volume} \ref{subsec:Impact of Speaker Distance}, \ref{subsec:Impact500hz}, and \ref{subsec:Impact of Speaker's Orientation}). This situation exists in numerous scenarios (see Sections \ref{sec:Threat Model} and \ref{sec:pilot study}), where DPSs are often installed in corridors, near entrance and exit doors, near diffusers, or within ventilation grilles, in industrial facilities. We run an experiment in the cleanroom corridor (see Section \ref{sec:pilot study}, Fig. \ref{fig:experimentesetup}, \& \ref{fig:DPSsclosetohuman})  to calculate what percentage of real-world conversations meet all attack constraints. In our case, the DPSs are located on the corridor wall near the entrance doors and far away from the entrance doors. A volunteer counts for 3 days on how many people (they are unaware of this experiment) are doing normal talking ($\sim$60 dB) while passing through that corridor with $\sim\leq$1.2 m close and $\sim0^0$ orientation. Table \ref{table:feasilibty} shows that the possibility of  a conversation satisfying all these constraints is higher near the entrance door compared to the non-entrance door locations. Our volunteer has observed that $\sim$25.3\% of 67 people near the entrance satisfied the attack constraints whereas this is 12.3\% for non-entrance locations. These are empirical findings and may change with DPS's locations.}

\begin{table}[ht!]
\vspace{-01.200em}
    \footnotesize
    \centering
    \renewcommand{\arraystretch}{0.85}
    \setlength{\tabcolsep}{03.50pt}
     \caption{\mycolor{N = Total people observed, S = Satisfying the constraints in \%.}} 
    \vspace{-01.00em}
    {\color{red}
    \begin{tabular}
     {c|c|c|c|c|c|c|c|c}
    \hline

          \cellcolor [gray]{0.85}\textbf{Location} & \multicolumn{2}{c|} {\cellcolor [gray]{0.85} \textbf{Day 1}}  & \multicolumn{2}{c|} {\cellcolor [gray]{0.85} \textbf{Day 2}} & \multicolumn{2}{c|} {\cellcolor [gray]{0.85} \textbf{Day 3}} & \multicolumn{2}{c} {\cellcolor [gray]{0.85} \textbf{Total}}\\ 
          \hline
             & \cellcolor [gray]{0.85} \textbf{\#N}   & \cellcolor [gray]{0.85} \textbf{S\%} & \cellcolor [gray]{0.85} \textbf{\#N} & \cellcolor [gray]{0.85} \textbf{S\%}  & \cellcolor [gray]{0.85} \textbf{\#N} & \cellcolor [gray]{0.85} \textbf{S\%} & \cellcolor [gray]{0.85} \textbf{\#N} & \cellcolor [gray]{0.85} \textbf{S\%}\\ 
         
        \hline
         Near entrance &  18 & 16.6\% &  28 & 35.7\% & 21 & 19.1\% & 67 & 25.3\%\\
        \hline
        Non-entrance &  15 & 6.7\% &  24 & 16.6\% & 18 & 11.2\% & 57 & 12.3\%\\
    \end{tabular}
    }
    \label{table:feasilibty}
    \vspace{-01.200em}
\end{table}




\vspace{-0.5em}
\section{Countermeasures}
\label{sec:Countermeasures}
\vspace{-0.2em}


\textbf{Audio damping:} The first method is to damp the audio  at the input of the pressure ports without affecting the normal pressure measurement. One solution is to use a pressure sampling tube longer than 1 m. Another solution is to enclose the pressure pickup device using a box-like enclosure filled with sound-damping foam to dampen the audio (see Fig. \ref{fig:defense} in Appendix \ref{appensec:defense}). \mycolor{Both defenses are cheap and easy to adopt.} 


\textbf{Reducing sampling frequency of DPSs:} Another intuitive solution is to keep the sampling frequency less than 500 Hz. However, reducing sampling frequency is not often possible in some specific applications, where small pressure changes must be detected accurately and quickly for a stable control process. Therefore, for these closely controlled applications, an audio damping device should be used instead of reducing the sampling frequency of pressure sensors.



\vspace{-0.5em}
\section{Related Work}
\label{sec:Related Work}
\vspace{-0.3em}

\textbf{Acoustic eavesdropping:} 
\mycolor{Extensive studies on different sensor modalities, such as} lasers \cite{muscatell1984laser,sami2020spying}, inertial measurement units (IMU) \cite{anand2018speechless,ba2020learning,han2017pitchln, michalevsky2014gyrophone,hu2022accear}, wireless signals \cite{hu2023mmecho,hu2022milliear, wang2022mmphone,wang2020uwhear,wei2015acoustic,xu2019waveear}, optical sensors \cite{long2023side,nassi2022lamphone}, vibration motors \cite{roy2016listening}, and hard drives \cite{kwong2019hard} \mycolor{are explored to reveal great threats to speech privacy. mmEcho \cite{hu2023mmecho}, mmEve \cite{wang2022mmeve}, and mmSpy \cite{basak2022mmspy} use mmWave sensors to capture vibrations and reconstruct speech. However, they cannot reconstruct speech with full intelligibility from the narrowband vibration data. IMU sensors }\cite{ba2020learning,han2017pitchln,michalevsky2014gyrophone}  can do digit inference, gender recognition, and limited hot-word reconstruction. However, these works still suffer from the narrowband condition of vibration-based side channels and can only recover band-limited sound with
damaged intelligibility. \mycolor{GlowWorm \cite{nassi2021glowworm} and Lamphone \cite{nassi2022lamphone} employ optical sensors to 
recover speech with limited intelligibility. However, they do not recover clean phases, and, as a result, they will not work in transient noisy conditions, like WaLi.}

Recently, AccEar \cite{hu2022accear} and VibSpeech \cite{wang2024vibspeech} revealed the possibility of using generative adversarial networks \mycolor{to reconstruct intelligible speech with unrestricted vocabulary. 
WaLi solves the following two problems of AccEar and VibSpeech:}

\textbf{1.} 
Compared to AccEar \cite{hu2022accear}, our WaLi does not require any \{audio, pressure sensor data\} pair from a specific target victim for training to recover intelligible audio. \mycolor{This makes our attack model more realistic, as it is not practical to have access to the victim's speech for training.}

\textbf{2.} 
WaLi can recover speech in the presence of transient noise injected from HVAC, whereas, VibSpeech \cite{wang2024vibspeech} has not been tested for such transient noise. Moreover, WaLi does not require SpkEnc-type encoders and extra vocoders like VibSpeech, as WaLi uses a complex-valued network to extract features from the magnitude and phase of the speech.

\textbf{Pressure sensor-based eavesdropping:} Recently, BaroVox \cite{achamyeleh2024fly} proposed to use DPSs to eavesdrop. The main differences between WaLi and BaroVox are: (1) BaroVox only recognizes hot words or phrases, whereas WaLi can reconstruct intelligible speech with unrestricted vocabulary. (2) BaroVox has not been tested for the transient noise condition. Therefore, it is not realistic for the real-world scenario in HVAC systems. 

\begin{table}[h!]
\vspace{-001.20em}
	\footnotesize
    \renewcommand{\arraystretch}{0.65}
	\centering
		\caption{Comparison between WaLi and recent work.}
		\vspace{-0.75300em}
		\label{table:strength of HALC}
		\begin{tabular}{|p{3.60cm}|p{2.6cm}|p{1.3cm}|}
			\hline
			Comparison  &  Recent works  & WaLi \\
			\specialrule{.15em}{0em}{0em}
             Attacking pressure sensors  & Only \cite{achamyeleh2024fly} does hot word recognition & Unrestricted vocabulary \\
			\hline
			 Unrestricted vocabulary   &\cite{muscatell1984laser,sami2020spying,anand2018speechless,ba2020learning,han2017pitchln, michalevsky2014gyrophone,hu2022accear,hu2023mmecho,hu2022milliear, wang2022mmphone,wang2020uwhear,wei2015acoustic,xu2019waveear,long2023side,nassi2022lamphone,roy2016listening,kwong2019hard,hu2023mmecho,wang2022mmeve,basak2022mmspy} cannot do \xmark & Can do \ding{51} \\
			\hline
   		Does not require \{audio, pressure data\} pair from a  target & AccEar \cite{hu2022accear} does not support \xmark  & Support \ding{51}\\
			\hline
			 Phase reconstruction in noise & \cite{wang2024vibspeech} cannot do \xmark & Can do \ding{51}\\
			 \hline
		
		\end{tabular}
\end{table}
\vspace{-01.00em}
\vspace{-0.5em}
\section{Conclusion}
\label{sec:Conclusion}
\vspace{-0.2em}

We expose a new speech \mycolor{threat that can recover} intelligible speech up to 8 kHz from severely aliased pressure sensor data, having a sampling frequency greater than 500 Hz. If \mycolor{a victim unknowingly comes} close to the pressure ports and continues a private
speech, the attacker can eavesdrop on the victim using our WaLi. WaLi can reconstruct audio even if the speaker is different and is not seen by the model during training time.  \mycolor{This makes WaLi more flexible compared} to previous work. Using our WaLi, an attacker can secretly listen to natural conversation behind the wall that is the least unexpected. Moreover, we comprehensively evaluate WaLi using six metrics that have not been done before. 

\Urlmuskip=0mu plus 1mu\relax
\bibliographystyle{IEEEtran}

\bibliography{v1}

\appendices

\vspace{-0.0em}
\section{Pitch, Phonemes, and Intelligible Bandwidth}
\label{append:Speech Production}
\vspace{-0.0em}

The human vocal folds vibrate at different frequencies. The strongest and slowest vibration is the pitch, and the faster vibrations that occur simultaneously are called harmonics. The pitch and harmonics generated by the vocal folds are selectively converted into phonemes \cite{fant1960acoustic}.  

Phonemes are the basic sound units of speech \cite{fant1960acoustic}. The frequencies in the phonemes are distributed particularly in terms of formants, pitches, and harmonics. For example, English has 44 phonemes, and the vowel phonemes of English have a frequency range of $\sim$4000 Hz, where the first few formants carry a significant amount of energy. The stops phonemes have a broadband energy often above 1000 Hz with an aperiodic burst. The fricative phonemes have energy concentrated in 4-8 kHz. Therefore, to capture intelligible speech with all phonemes, the sampling frequency should be at least 16 kHz (twice the maximum frequency content, 8 kHz). However, vowel phonemes and critical formants fall within the 300-4000 Hz range. Therefore, for speech intelligibility, only 8 kHz sampling is sufficient, as up to 4000 Hz bandwidth is enough for intelligible speech reconstruction.

\begin{table}[ht!]
\vspace{-0.9800em}
\footnotesize
    \centering
    \caption{Pressure ranges for different voice conditions \cite{ccohs2019soundpressure}.}
    \vspace{-0.7800em}
    \begin{tabular}{l | l|l}
    \hline
        \cellcolor [gray]{0.85}\textbf{Condition} & \cellcolor [gray]{0.85}\textbf{in dB} & \cellcolor [gray]{0.85}\textbf{in Pa} \\ 
        \hline
        \hline
        Quiet whisper & 40 dB & 0.02 Pa  \\ 
        \hline
       Conversation & 60 dB & 0.2 Pa \\
       \hline
        Loud speech & 70-100 dB & 0.6-10 Pa \\ 
        \hline
    \end{tabular}

    \vspace{-01.10em}
    \label{table:voice SPL}
\end{table}

\vspace{-0.3em}
\section{Impact of Speech on Pressure Sensors}
\label{append:Impact of Speech on Pressure Sensors}
\vspace{-0.2em}

Table \ref{table:pressureHVACs} indicates that DPSs have a sampling frequency of 0.5-2 kHz. However, at least 8 kHz sampling is required to recover intelligible speech. Therefore, a natural question is: \textit{Will it be possible to capture intelligible speech with at least 0.5 kHz sampling frequency using pressure sensors?}. The pitch frequencies for a male speaker vary between 85-180 Hz, whereas for a female speaker, they vary between 165-255 Hz \cite{kent1997speech}. Although the 0.5 kHz sampling rate can capture pitches and a few harmonics,  it is not sufficient to provide perfect intelligibility \cite{fant1960acoustic}, as most harmonics at high frequencies will be missing from the bandwidth. 

\begin{figure}[h]
\vspace{-0.96320em}
\centering
\includegraphics[width=0.48\textwidth,height=0.12\textheight]{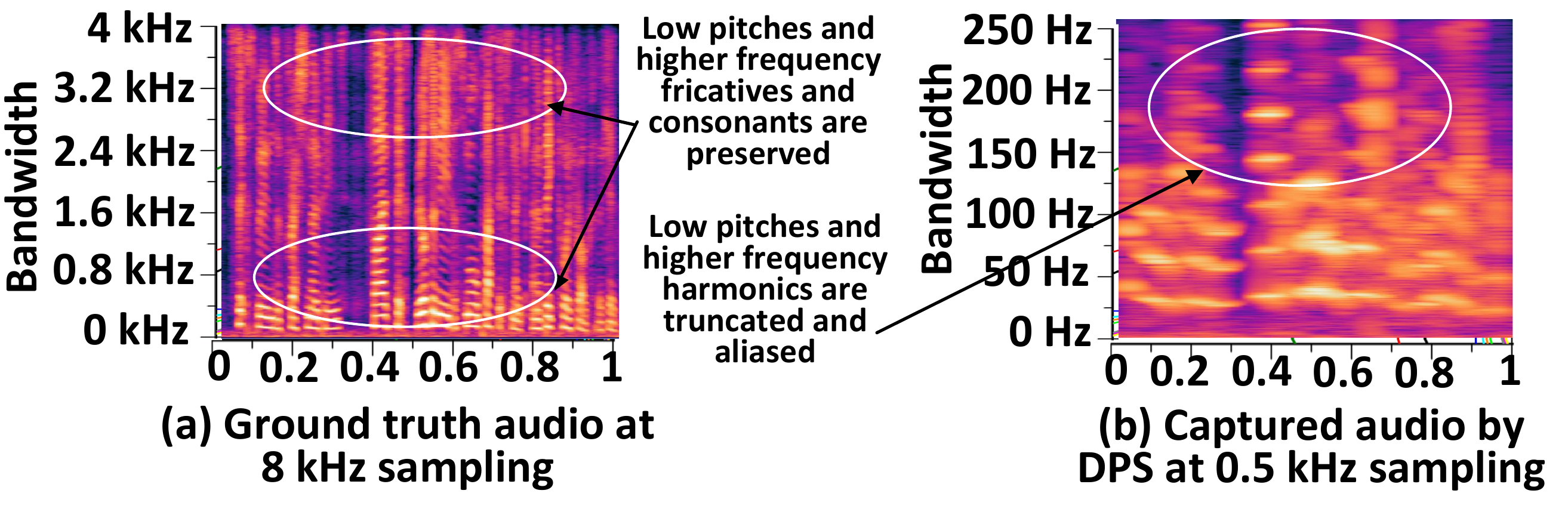}
\vspace{-01.18820em}
\caption{(Left) Low pitch and high frequency consonants are preserved. (Right) High-frequency fricatives get severely aliased.}
\label{fig:aliasedband}
\vspace{-0.500em}
\end{figure}

Fig. \ref{fig:aliasedband} shows the importance of various frequencies in the intelligibility of speech signals. Evidently, the higher frequency components that involve the use of fricatives and other consonants are critical for higher intelligibility (see Fig. \ref{fig:aliasedband} (Left)). Unfortunately, with a sampling rate of 0.5 kHz, high-frequency speech components are severely aliased in pressure sensors (see Fig. \ref{fig:aliasedband} (Right)).  WaLi provides a means to eavesdrop speech in full intelligibility by enhancing the severely aliased signals from pressure sensors in HVACs. This will seriously hamper the confidentiality of safety-critical systems, as nowadays most critical infrastructures have some form of HVAC in their design.


\section{Sensor Details}
\label{appensec:5SensorDetails}
\vspace{-0.1em}

\begin{table}[ht!]
\vspace{-0.500em}
    \footnotesize
    \centering
     \caption{Sensor details.} 
    \vspace{-01.00em}
    {\color{red}
    \begin{tabular}
     {c|c|c}
    \hline
          \cellcolor [gray]{0.85}\textbf{Part name} & \cellcolor [gray]{0.85} \textbf{Vendor}   & \cellcolor [gray]{0.85} \textbf{Range}\\
        \hline
         SDP810-125PA &  Sensiron  &  -125 - +125 Pa \\
         SDP810-500PA & Sensiron   &  -500 - +500 Pa \\
         SETRA264 &  Setra Systems &  -0.1 inch WC to +0.1 inch WC \\

 \end{tabular}
 }
    \label{appentable:sensor}
    \vspace{-0.200em}
\end{table}

\vspace{-0.3em}
\section{Details on Comprehensive Evaluation Metrics}
\label{append:Evaluation_Metrics}
\vspace{-0.1em}


We evaluate WaLi using the \mycolor{following six different metrics.}


\textbf{a) Log-Spectral Distance (LSD):} LSD measures the difference between the log-magnitude spectra of a ground truth signal and a reconstructed one from the pressure data.


\textbf{b) Non-Intrusive Speech Quality Assessment - Mean Opinion Score (NISQA-MOS):} NISQA-MOS is used to estimate the \mycolor{perceived quality of reconstructed audio without needing a reference signal and gives values between 0 to 5.} 

\textbf{c) Scale-Invariant Signal-to-Distortion Ratio (SI-SDR):} SI-SDR measures how close a reconstructed signal is to a clean target signal, ignoring any difference in gain.

\textbf{d) Perceptual Evaluation of Speech Quality (PESQ):} PESQ is widely used to evaluate how a reconstructed signal compares to a reference (clean) signal, closely modeling human perception. PESQ has values between -0.5 to 4.5

\textbf{e) Short-Time Objective Intelligibility (STOI):} \mycolor{STOI predicts how understandable speech is to human listeners in the presence of noise and gives values between 0 to 1.}

\mycolor{\textbf{f) Word Error Rate (WER):} WER measures how many words are incorrectly reconstructed compared to a reference transcript \cite{morris2004wer}. A lower WER means better speech quality.}

\begin{figure}[h]
\vspace{-0.95320em}
\centering
\includegraphics[width=0.35\textwidth,height=0.1\textheight]{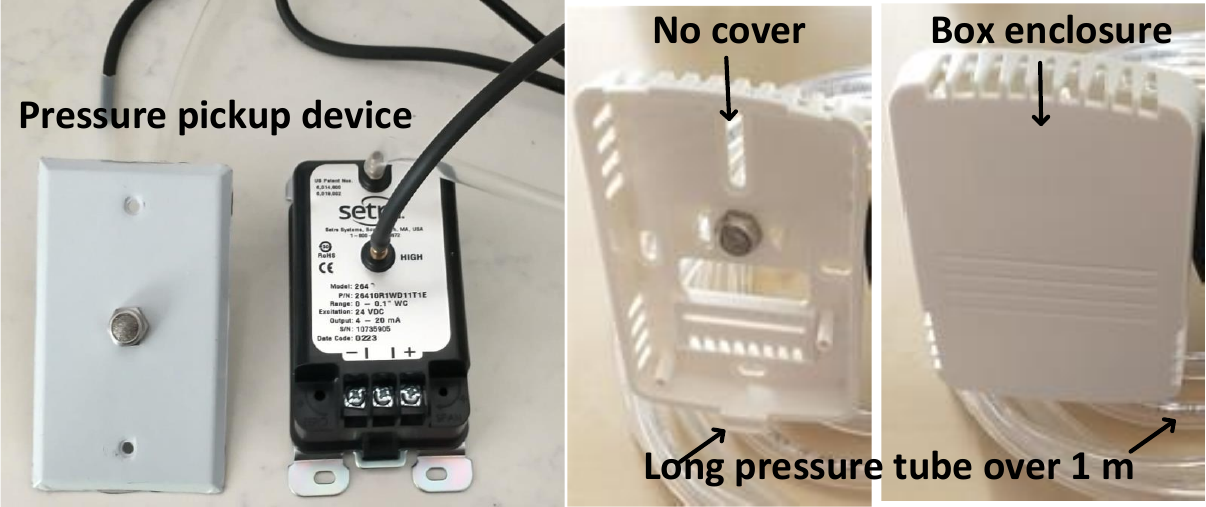}
\vspace{-0.90em}
\caption{Long pressure tube and box-like enclosure.}
\label{fig:defense}
\vspace{-0.8500em}
\end{figure}

\vspace{-0.3em}
\section{Audio damping}
\label{appensec:defense}
\vspace{-0.1em}

\mycolor{
The first method (see Fig. \ref{fig:defense}) is to damp the audio using a damping device at the input of the pressure ports without affecting the normal pressure measurement. One solution is to use a pressure sampling tube longer than 1 m to dampen the audio significantly. Another solution is to enclose the pressure pickup device using a box-like enclosure filled with sound-damping foam to dampen the audio. Both of these countermeasures are cheap and easy to adopt.}

\vspace{-0.0em}
\section{Comparative Analysis with Recent Work}
\label{appensec:comanalysis}
\vspace{-0.1em}

\begin{table}[h!]
	\footnotesize
	\centering
		\caption{Comparison between WaLi and recent work.}
		\vspace{-0.75300em}
		\label{table:strength of HALC}
		\begin{tabular}{|p{3.60cm}|p{2.6cm}|p{1.3cm}|}
			\hline
			Comparison  &  Recent works  & WaLi \\
			\specialrule{.15em}{0em}{0em}
             Attacking pressure sensors  & Only \cite{achamyeleh2024fly} does hot word recognition & Unrestricted vocabulary \\
			\hline
			 Unrestricted vocabulary   &\cite{muscatell1984laser,sami2020spying,anand2018speechless,ba2020learning,han2017pitchln, michalevsky2014gyrophone,hu2022accear,hu2023mmecho,hu2022milliear, wang2022mmphone,wang2020uwhear,wei2015acoustic,xu2019waveear,long2023side,nassi2022lamphone,roy2016listening,kwong2019hard,hu2023mmecho,wang2022mmeve,basak2022mmspy} cannot do \xmark & Can do \ding{51} \\
			\hline
   		Does not require \{audio, pressure data\} pair from a  target & AccEar \cite{hu2022accear} does not support \xmark  & Support \ding{51}\\
			\hline
			 Phase reconstruction in noise & \cite{wang2024vibspeech} cannot do \xmark & Can do \ding{51}\\
			 \hline
		
		\end{tabular}
\end{table}
\vspace{-01.00em}

\section{Sampling For Less than 500 Hz}
\label{appensec:500hzsampling}
\vspace{-0.1em}

\begin{figure}[h]
\vspace{-0.75320em}
\centering
\includegraphics[width=0.4\textwidth,height=0.1\textheight]{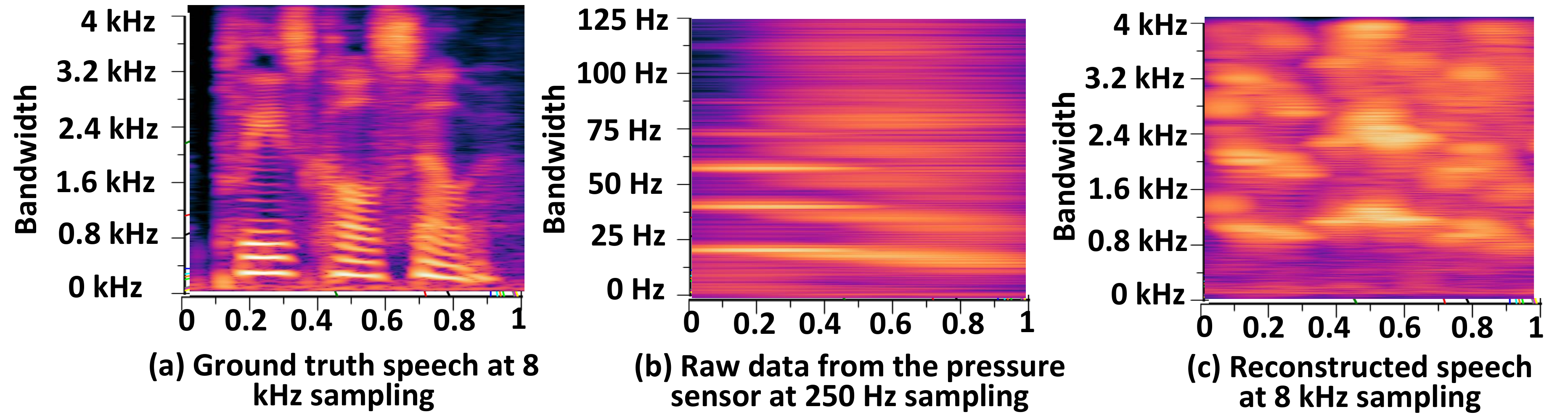}
\vspace{-0.90em}
\caption{Reconstructed speech at (c) from 250 Hz sampling frequency at (b) has severely degraded intelligibility.}
\label{fig:250hz}
\vspace{-0.7500em}
\end{figure}

\begin{figure}[ht!]
  \centering
\includegraphics[width=0.49\textwidth,height=0.32\textheight]{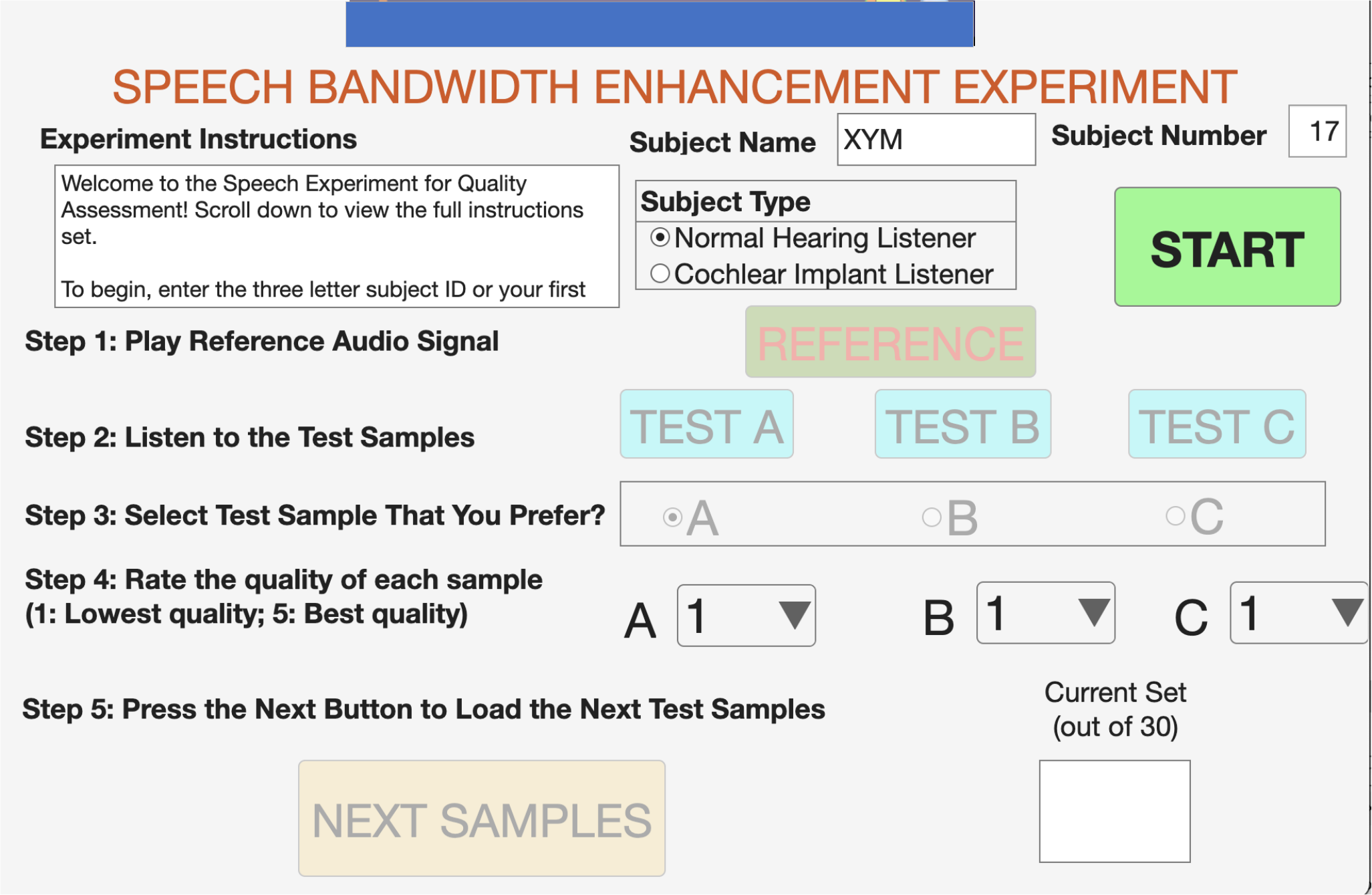}
\vspace{-0.750em}
  \caption{The MATLAB interface used in Subjective Tests.}
  \label{fig:subj_eval_interface_ss}
  \vspace{-01.60em}
\end{figure}

\section{Subjective Evaluation Details} 
\label{appensubsec:subj_eval_details}

\mycolor{To evaluate the performance of the proposed WaLi, a formal listening test has been conducted. A total of 18 undergraduate students with mixed demography who self-reported normal-hearing (NH) participants—comprising ten males and eight females with an average age of 22—participated in the study. The participants voluntarily join to rate the audios without any compensation. At first they are trained on how to assign scores based on perceived perceptual quality of audios. They are also briefed about the purpose of the experiments, potential risks, and about the outcome of this paper. All participants were native English speakers and used soundproof headsets to ensure consistent and distraction-free listening conditions. Each participant evaluated 30 sets of speech samples. Every set included two randomly presented versions of the same utterance: (i) the unprocessed pressure sensor signal, and (ii) the speech processed by the proposed WaLi.  Participants rated the perceptual quality of each sample on a 5-point Mean Opinion Score (MOS) scale, where 1 indicates the lowest and 5 the highest quality. The test used speech clips, which were processed under two different frequency band conditions: 500 Hz to 8 KHz and 2 kHz to 8 kHz. In the Fig. \ref{fig:subj_eval_interface_ss}, we have presented the MATLAB interface that we use to conduct the subjective evaluation.}

s


\end{document}